\documentclass[fleqn,usenatbib]{mnras}

\usepackage{newtxtext,newtxmath}

\usepackage[T1]{fontenc}

\DeclareRobustCommand{\VAN}[3]{#2}
\let\VANthebibliography\thebibliography
\def\thebibliography{\DeclareRobustCommand{\VAN}[3]{##3}\VANthebibliography}

\usepackage{graphicx}	
\usepackage{amsmath}	
\usepackage{hyperref}
\usepackage{xcolor}
\usepackage [latin1]{inputenc}
\usepackage{multicol}

	\usepackage{xspace}
	
	\newcommand{\eq}[1]{Eq.~(\ref{eq:#1})\xspace}

	\newcommand{\AU}{ \  \rm AU}
	
	\newcommand{\Ms}{ \   \rm M_\odot }
	\newcommand{\Msyr}{ \ \rm M_\odot \,  \rm yr^{-1} }

	\newcommand{\Me}{ \  M_\oplus}

	\newcommand{\Omegak}{\Omega_{\rm K}}

	\newcommand{\alphag}{\alpha_{\rm g}}
	\newcommand{\alphat}{\alpha_{\rm t}}

\title[]{The impact of pre-main sequence stellar luminosity on giant planet formation}

\author[H. F. Johnston et al.]{
Heather F. Johnston$^{1, 2}$\thanks{E-mail: h.johnston2@exeter.ac.uk (UoE)},
Olja Pani\'c$^{2}$,
Beibei Liu $^{3,4}$,
Patryk Jankowski $^{2}$
\\
$^{1}$ Department of Physics and Astronomy, University of Exeter, Stocker Road, Exeter, EX4 4QL, UK\\ 
$^{2}$School of Physics and Astronomy , University of Leeds, Woodhouse, Leeds LS2 9JT, England\\
$^{3}$Astronomical Institute, School of Physics, Zhejiang University, 38 Zheda Road, Hangzhou, 310027 China\\
$^{4}$Center for Cosmology and Computational Astrophysics, Institute for Advanced Study in Physics, Zhejiang University, Hangzhou, 310027 China\\
}

\date{Accepted XXX. Received YYY; in original form ZZZ}

\pubyear{2025}

\begin{document}
\label{firstpage}
\pagerange{\pageref{firstpage}--\pageref{lastpage}}
\maketitle

\begin{abstract}
Luminosities of pre-main sequence stars evolve during the protoplanetary disc lifetime.  This has a significant impact on the heating of their surrounding protoplanetary disks, the natal environments of planets. Moreover, stars of different masses evolve differently. However, this is rarely accounted for in planet formation models.
We carry out pebble-driven core accretion planet formation modelling with focus on the impact of pre-main sequence stellar luminosity evolution on giant planet formation around host stars in the range of $1{-}2.4\ \rm M_{\odot}$. We find that giant planet formation is sensitive to the evolution of stellar luminosity, specifically the locations and times at which giant planet formation can occur depend on it.  High stellar luminosity causes an increase in the scale height of the gas and pebbles, which may decrease the efficiency of pebble accretion, making it more challenging to form giant planets.  This has important consequences for the composition of these giant planets, stressing the need to incorporate such aspects into planet formation models.
\end{abstract}




\section{Introduction}

Giant planets contain the majority of mass and angular momentum within a planetary system making them instrumental in shaping planetary architecture.  It is also thought that they may play some deciding role in the habitability of terrestrial planets in planetary systems.  Thus, it is important to study their formation and evolution in order to understand planetary dynamics in systems like our own Solar System.  

The giant planet occurrence rate, $\eta_{\rm J}$, has been found to increase with stellar metallicity \citep{Gonzalez1997TheConnection, Santos2001ThePlanets, Santos2004SpectroscopicFormation, Fischer2005THE1, Udry2007StatisticalExoplanets, Johnson2010GIANTPLANE, Sousa2011SpectroscopicCorrelation, Mortier2012TheStars, Mortier2013OnCorrelation}. This correlation is commonly attributed to the core accretion model of planet formation \citep{Liu2020APlanets, Drazkowska2023PlanetExoplanets} (more metals present allows the planetary core to build more efficiently, allowing planets to undergo runaway gas accretion at earlier times where there is likely to be a greater abundance of gas in the disk).  

Nonetheless, giant planet(s) are relatively rare and are only found around $10{-}20\%$ of stars \citep{Cumming2008ThePlanets, Johnson2010GIANTPLANE, Mayor2011ThePlanets}.  $\eta_{\rm J}$ peaks around stars of masses $M_{\star} \sim 1.7\ M_{\odot}$ \citep{Reffert2015PreciseMetallicity, Wolthoff2022PreciseSurveys}. 
 The exact cause of this $M_{\star}$-dependence of  $\eta_{\rm J}$ is still unclear. However, we know host stars have major influence on their surrounding protoplanetary disk, specifically dictating the lifetime of the disk.  Disk lifetime is one of the most important parameters to consider when studying giant planet formation via core accretion as it must build the planetary core at a fast enough rate to surpass isolation mass and accrete gas while there is still material in the environment to be able to do so. 
 Hence, pre-main sequence (PMS) luminosity evolution of the host star plays a role in controlling the rate of disk evolution which, in turn, controls what kind of planets are able to form. 

\cite{Payne2007TheDwarfs} studied the impact of stellar luminosity evolution in dust grain disks around young, brown dwarfs $(M_{\star}{=}0.05 \Ms)$.  They found that the inclusion of evolving stellar luminosity affected (i) the gas temperature and (ii) the location of the water snowline (where $a_{\rm ice}{=}2.7(L_{\star} / L_{\odot})^{1/2}$ [AU], from \cite{Hayashi1981StructureNebula}).  This had a relatively small impact on their planet formation results as there is little or no gas accretion onto protoplanets around brown dwarfs \citep{Payne2007TheDwarfs}.

The luminosity evolutionary tracks of PMS intermediate-mass stars differ significantly from their low- or solar-mass equivalents.  Low-mass stars decrease in luminosity as they evolve towards the main sequence.  This means there is a decreased amount of incident irradiating flux onto the disk, resulting in the disk temperature cooling.  In contrast, the stellar luminosity of stars $\geq 1.5 \Ms$ increases as they evolve towards the main sequence, causing the temperature in their surrounding disks to be warmer at timescales of just a few Myrs \citep{Miley2020TheDiscs}.  A fully convective star on its Hayashi track will shrink due to the loss of radiative energy (Kelvin-Helmholtz contraction) \citep{Hayashi1961StellarContraction}.  This loss of radiative energy causes an increase in the internal temperature with time (virial theorem) and decrease in the stellar internal opacity with time (Kramers law) \citep{Kunitomo2021PhotoevaporativeStars}.  Simultaneously, the star is becoming less luminous while mostly maintaining the same stellar surface temperature. 
 Then, either a radiative core is developed (stellar luminosity begins to increase over time) and the star leaves the Hayashi track, or it joins the main sequence as nuclear fusion begins.  More massive stars ($M_{\star}{>}1.5 \Ms$) have a shorter Kelvin-Helmholtz timescale and develop their radiative core and hotter photosphere at a faster rate than less massive stars \citep{Palla1993TheStars, Siess2000AnStars, Kunitomo2021PhotoevaporativeStars}.  
Before the main sequence and becoming more luminous, these intermediate-mass stars do experience an initial drop in luminosity, it is not as significant a decrease as seen in low-mass stars and the luminosity increases again within just a few Myr \citep{Palla1993TheStars, Siess2000AnStars}.  Due to this inherent divergence in evolution between low- and intermediate-mass stars, \cite{Panic2017EffectsLocation} infer different physical and chemical conditions in the disks around low- and intermediate-mass stars.  As a consequence, the irradiated temperature of the disk evolves alongside the stellar luminosity affecting physical processes in disks and disk chemistry \citep{Panic2017EffectsLocation,Miley2020TheDiscs}.   However, the impact of this evolution on planet formation remains unexplored.  While an evolving stellar luminosity is not unprecendeted in planet formation modelling \textit{e.g.} \citep{Emsenhuber2021TheResults, Kessler2023TheFormation}, its impact is rarely discussed in detail.

Models employed in this work builds on the previous ones by \cite{Johnston2024FormationStars}, where we carried out pebble-driven core accretion planet formation modelling to investigate the trends and optimal conditions for giant planet formation around host stars in the range of $1{-}2.4 \Ms$.  The stellar luminosity parameter scaled with stellar mass $(L_{\star} \propto M_{\star}^{1{-}4})$ but was otherwise fixed.  
  In that study, we found that giant planets are most likely to form in systems that have a large initial disk radii; higher initial accretion rates; pebbles approximately a millimetre in size; and embryos with moderate birth locations of $\sim 10$ AU.  

In this paper we explore the impact of stellar luminosity on giant planet formation.  We use self consistent temperature profiles calculated using stellar parameters at various stages throughout the PMS evolution process for the stellar mass range of $1{-}2.4 \Ms$ \citep{Siess2000AnStars, Min2009RadiativeDisks}.  We use the pebble-driven core accretion model from \citep{Liu2019Super-EarthMasses}, as adapted in \citep{ Johnston2024FormationStars}, now including the stellar-mass dependent and time evolving $L_{\star}$.  We seek to identify the key physical properties behind giant planet formation that are impacted by stellar luminosity.  Since the aim of this paper is purely comparative and focusing on stellar luminosity alone, we fix other parameters to be the same as in \cite{Johnston2024FormationStars}.  This paper is organised as follows.  The model is described in Section \ref{sec:method}.  We investigate the final planetary masses, migration and composition in Sections \ref{sec:gromap} and examine the impact of stellar luminosity on giant planet formation in Section \ref{sec:comp}.  
  Our findings are then summarised in Section \ref{sec:conc}. 


\section{Method}
\label{sec:method}
We employ the planet formation model of \cite{Liu2019Super-EarthMasses} and recapitulate the key equations in this section.  Readers are recommended to  go through their Section 2 for details.  The disk condition presented here is adapted to the circumstance around host stars of masses $1{-}2.4\Ms$, as in \cite{Johnston2024FormationStars}, with the addition of evolving stellar luminosity. This, in turn, implies improved midplane temperatures, and CO and $\rm CO_{2}$ snowline locations for our model.   


\subsection{Disk model}
\label{sec:disk}

We adopt a conventional $\alpha$ viscosity prescription for the gas disk angular momentum transport \citep{Shakura1973BlackAppearance}.  For simplicity and ease of isolating the heating affects, we choose a viscously smooth disk without any structural discontinuities for simplicity. Note that the presence of gaps and rings in the disk caused by giant planet(s) or (magneto)-hydrodynamical instabilities  can substantially impact the growth and migration of protoplanets.  The disk accretion and surface density can be linked through $\dot M_{\rm g}{=}3 \pi \nu \Sigma_{\rm g}$ with viscosity $\nu{=}\alphag c_{\rm s} H_{\rm g}$, where  $c_{\rm s}{=}H_{\rm g} \Omegak$ is the sound speed, $H_{\rm g}$ is the gas disk scale height and $\Omegak$ is the Keplerian frequency. The dimensionless viscous coefficient $\alphag$ determines the global disk evolution and gas surface density.  We set it to be a fixed value of $10^{-2}$, the same as \cite{Liu2019Super-EarthMasses, Johnston2024FormationStars}.

We consider the disk angular momentum transport by gas accretion/spreading from the internal viscous stress \citep{Lynden-Bell1974THEVARIABLES} and evaporation by high energy UV and X-ray photons emitted from the central star \citep{Alexander2013TheDisks}. The evolution of the disk accretion rate is expressed as
  \begin{equation}
 \dot{M}_{\rm g} = \begin{cases}
 {\displaystyle   \dot{M}_{\rm g0} \left[ 1 + \frac{t}{\tau_{\rm vis}} \right] ^{-\gamma} }, 
 \hfill  \ t{<}t_{\rm pho}, \\
 {\displaystyle \dot{M}_{\rm g0} \left[ 1 + \frac{t}{\tau_{\rm vis}} \right]^{-\gamma} \exp \left[ - \frac{t - t_{\rm pho}}{\tau_{\rm pho}} \right], }  \hfill \   t{\geq}t_{\rm pho},\\
\end{cases}
\label{eq:mdotg}
\end{equation}
where  $\dot M_{\rm g0}$ is the initial disk accretion rate; the $t - t_{\rm pho}$ term refers to the mass loss due to stellar photoevaporation; $\gamma =( 5/2+s) / (2+s)$, and $s$ is the gas surface density gradient ($s=-3/8$ in viscous-dominated regions of the disk; $s=-15/14$ in irradiated-dominated regions of the disk).  $t_{\textrm{pho}}$ is defined as the time when $M_{\textrm{g}}$ falls below $\dot{M}_{\textrm{pho}}$, marking the onset of photoevaporation-dominated disk dispersal.  
$\tau_{\rm vis}$ and $\tau_{\rm pho}$ are the characteristic viscous accretion and photoevaporation timescales from Eqs. (3) and (7) of \cite{Liu2019Super-EarthMasses}:

\begin{equation}
\label{eq:tvis}
    \tau_{\rm vis}{=} \frac{1}{3(2 + s)^{2}} \frac{r^{2}_{\rm d0}}{\nu_{0}}.
\end{equation}

$r^{2}_{\rm d0}$ and $\nu_{0}$ are the initial characteristic size of the gas disk and the viscosity at that radius.  Notably, the viscosity is dependent on sound speed $(c_{\rm s})$, scale height of the gas disk $(H_{\rm g})$, and Keplerian angular frequency $(\Omegak)$, through $\nu_{0}{=}\alpha c_{\rm s} H_{\rm g}$.  These are intrinsically dependent on the mass of the host star and irradiated temperature of the disk (and hence the luminosity of the host star, $L_{\star}$).  The viscous timescale, $\tau_{\textrm{vis}}$, is recalculated at each timestep based on disc properties.   The timescale for a disk to adjust viscously is on the order of the local viscous time, $\tau_{\textrm{vis}} \sim r^{2} / \nu $.  For the inner disk regions ($r \lesssim 10$ AU), $\tau_{\textrm{vis}}$ is typically short ($\sim 10^{4}{-}10^{5}$ years) compared to the Myr-timescale of significant stellar evolution of a $2.4 \textrm{M}_{\odot}$ star.  Therefore, the assumption of a quasi-steady state at each timestep is a reasonable approximation.  The authors note that one key improvement of this implementation for future work would be via some constant subtraction of the mass-loss via photoevaporation term.

The gas flux changes sign at: 

\begin{equation}
    \label{eq:rc}
    r_{\textrm{c}} = r_{\textrm{d0}} \left[ \frac{1 + t / \tau_{\textrm{vis}}}{2(2+s)}\right]^{1/(2+s)}.
\end{equation}

The gas disk mass $M_{\rm d}(t) = \int_{0}^{r_{\rm d}} 2 \pi r \Sigma_{\rm g} (r, t)\textrm{d}r$.  Throughout our study, the disk mass is inferred to be correlated with the mass of the host star, \textit{e.g.} $M_{\textrm{d}} \propto M_{\star}$.   We note that while some dust mass estimates suggest a steeper-than-linear scaling \citep{Pascucci2016ARELATION, Manara2016EvidenceDisks}, these may not directly reflect he gas properties due to the faster evolution timescales of dust.  Since our disk evolves viscously, the mass accretion rate onto the central star $(\dot{M}_{\rm g})$ is directly correlated with the mass of the disk $(M_{\rm d})$ \citep{Hartmann1998ACCRETIONNURIA}. Thus, the ratio between $\dot{M}_{\rm g}$ and $M_{\rm d}$ relates to the viscous timescale at the outer edge of the disk (Equation \ref{eq:tvis}).  Observationally, \citep{Manara2016EvidenceDisks} determined accretion rates from UV excess with VLT/X-shooter spectrographs and disk masses from sub-mm continuum and CO line emission with ALMA.  They established a statistically significant correlation between $\dot{M}_{\rm g}$ and $M_{\rm d}$ with a slope slightly less than one \citep{Manara2016EvidenceDisks}.

 We specifically account for the X-ray driven photoevaporation, and the critical mass-loss rate is given by \citep{Owen2012TwoDiscs}: 
 \begin{equation}
 \begin{split}
   \frac{ \dot{M}_{\rm pho}}{ M_\odot \rm yr^{-1}}  = 6 \times 10^{-9} \left( \frac{M_{\star}}{1 \ M_\odot} \right)^{-0.07} \left( \frac{L_{X}}{30 \rm \ ergs^{-1}} \right)^{1.14}  \\
    {=} 3 \times 10^{-8} \left( \frac{M_{\star}}{2.4 \ M_\odot} \right)^{1.6}.
    \end{split}
    \label{eq:pho}
  \end{equation}

The latter equality in Equation \ref{eq:pho} is derived from the stellar X-ray luminosity and mass relation $L_{\textrm{X}} \propto M_{\star}^{1.5}$ from \cite{Preibisch2005THESTARS}, which is consistent with the results of \cite{Gudel2007TheXEST}.  We fix $L_{\textrm{X}}$ while varying $L_{\star}$.  In reality, $L_{\textrm{X}}$ also evolves with time \citep{Kunitomo2021PhotoevaporativeStars}.  A coupled, evolving X-ray luminosity would affect the photoevaporation rate (Equation \ref{eq:pho} and thus disk lifetime.  Exploring this coupled evolution is beyond the scope of this study but represents an interesting future avenue, particularly for more massive stars where a shift from predominately X-ray photoevaporation to FUV occurs.

\begin{figure}
\centering
\includegraphics[width = \linewidth]
{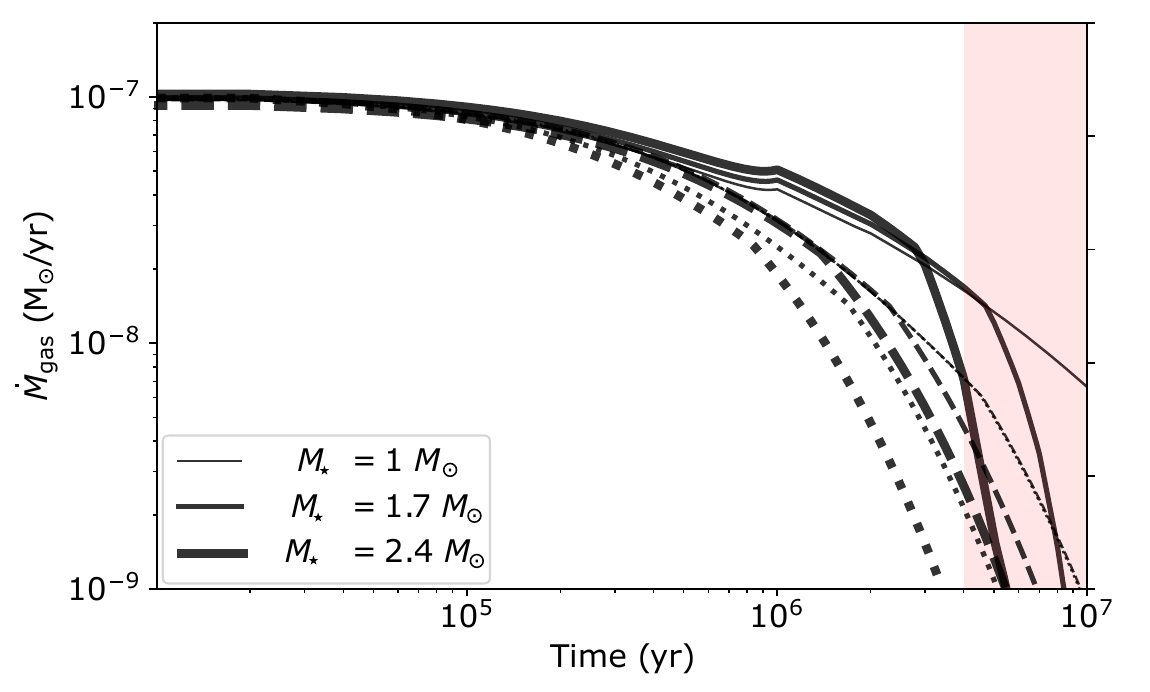}
\caption{Evolution of disk accretion rate.  When $\dot{M}_{\rm g} < \dot{M}_{\rm pho}$ (Equation \ref{eq:mdotg}, the disk angular momentum is transported by viscous accretion and later the gas removal is driven by stellar X-ray photoevaporation.  Three systems with the same initial disk accretion rates $(10^{-7} \Ms \rm yr^{-1})$ and initial disk size $(R_{\rm d0}{=}120$ AU) are shown around different masses of stars $(M_{\star}{=}1 \Ms$, $1.7 \Ms$, and $2.4 \Ms)$. The dashed and dotted lines show the fixed luminosity approximations for 1, 1.7, and $2.4 \Ms$ stars ($L_{\star}{=}M_{\star}^{\beta}$) when $\beta{=}2$ and $\beta{=}4$, respectively. The pink box indicates the regime where the $2.4 \Ms$ star becomes radiative and FUV becomes the dominant photoevaporation mechanism.  The small increase or 'kink' visible at $t \sim t_{\textrm{pho}}$ is an artifact of the transition between the viscous evolution and photoeavporation-dominated evolution formulae in Equation \ref{eq:mdotg}.  }  
\label{fig:acc_ev}
\end{figure}

Figure \ref{fig:acc_ev} shows the same initial accretion rate $(10^{-7} \Ms \rm yr^{-1})$ for the same initial disk size $(R_{\rm d0}{=}120$ AU) around three different host star masses $(M_{\star}{=}1 \Ms$, $1.7 \Ms$, and $2.4 \Ms)$, respectively.  The disk accretion rate declines as the disk evolves viscously ($t < t_{\rm pho}$ in Equation \ref{eq:mdotg}) and there is a significant drop in latter stages caused by stellar photoevaporation ($t \geq t_{\rm pho}$ in Equation \ref{eq:mdotg}).  The disks around more massive stars remain at a fractionally higher accretion rate but evolve faster due to the threshold for the onset of photoevaporation, $t_{\rm pho}$, scaling with $M_{\star}$, as detailed above.  Thus, X-ray photoevaporation becomes the dominant mass-loss mechanism at a faster rate around more massive stars resulting in the reduced disk lifetimes as $M_{\star}$ increases.  The time for the onset of X-ray photoevaporation ($t_{\rm pho}$) is shown in Table \ref{tab:tpho} for the stellar masses and luminosity prescriptions detailed in Figure \ref{fig:acc_ev}.  

\begin{table}
    \centering
    \begin{tabular}{|c|c|c|}
    \hline
    \hline
       $M_{\star}$ $[\Ms]$ & $L_{\star}$-prescription & $t_{\rm pho}$ onset [Myr] \\
       \hline 
        1 & $L_{\star} \propto M_{\star}^{2}$ & 4.5 \\
        1.7 & $L_{\star} \propto M_{\star}^{2}$ & 2.4 \\
        2.4 & $L_{\star} \propto M_{\star}^{2}$ & 1.4 \\
        1 & $L_{\star} \propto M_{\star}^{4}$ & 4.5 \\
        1.7 & $L_{\star} \propto M_{\star}^{4}$ & 1.75 \\
        2.4 & $L_{\star} \propto M_{\star}^{4}$ & 0.85 \\
        1 & PMS-evolving $L_{\star}$ & 8 \\
        1.7 & PMS-evolving $L_{\star}$ & 3 \\
        2.4 & PMS-evolving $L_{\star}$ & 1.7 \\
    \hline
    \hline 
    \end{tabular}
    \caption{Times for the onset of photoevaporation $t_{\rm pho}$ for the stellar masses and luminosity prescriptions in Figure \ref{fig:acc_ev}, as defined in Equation \ref{eq:pho}.  }
    \label{tab:tpho}
\end{table}
  

It is worth mentioning that \cite{Kunitomo2021PhotoevaporativeStars} examined the long-term disk evolution around stars of $0.5{-}5\ M_\odot$  and found that the X-ray luminosity of more massive stars decreases while the FUV luminosity rapidly increases $({\sim} 1$ Myr for stars ${\gtrsim} 3 \ M_\odot$).  The critical mass for this FUV-dominated photoevaporation mass-loss is thought to be around $2.5 \ M_\odot$ when the Kelvin-Helmholtz timescale is comparable with the disk dispersal timescale \citep{Kunitomo2021PhotoevaporativeStars}.  Thus, the X-ray dominated photoevaporation regime that we investigate in this work with Equation \ref{eq:pho} should be restricted to the first 8 Myr for $1.7 \Ms$ star and 4 Myr for $2.4 \Ms$ star to retain coherence with the latest literature.  This is noted by the pink box in Figure \ref{fig:acc_ev}.    We highlight this shift in dispersal mechanism where necessary in later sections.

We employ a two-component disk structure, including an inner viscously heated region and an outer stellar irradiated region. The gas surface density and disk aspect ratio are adopted from Eqs. (8)-(13) of \cite{Liu2019Super-EarthMasses} and $[\rm vis]$ refers to the inner viscously heated disk.  $[\rm irr]$ refers to the outer disk region heated entirely by stellar irradiation (optically thin):
  \begin{equation}
 \frac{\Sigma_{\rm g}}{ \rm g \ cm^{-2}}  =
  \begin{cases}
 {\displaystyle   465
    \left( \frac{\dot M_{\rm g}}{10^{-7} \Msyr} \right)^{1/2}  \left(\frac{M_{\star}}{2.4 \ M_{\odot}} \right)^{1/8} } \\
   {\displaystyle  \left(\frac{r}{1 \AU} \right)^{-3/8}  }
     \hfill  [\mbox{vis}],  \vspace{0.6cm}\\
 {\displaystyle  1613 \left( \frac{\dot M_{\rm g}}{10^{-7}  \Msyr} \right)
\left(\frac{M_{\star}}{2.4 \ M_{\odot}} \right)^{9/14}} \\
{\displaystyle  \left(\frac{L_{\star}}{2.4^4 \ L_{\odot}} \right)^{-2/7}  \left(\frac{r}{1 \AU} \right)^{-15/14} }
  \hfill  [\mbox{irr}], 
\end{cases}
\label{eq:sigma}
\end{equation}
and
  \begin{equation}
h_{\rm g} = \begin{cases}
 {\displaystyle   0.034
\left( \frac{\dot M_{\rm g}}{10^{-7} \Msyr} \right)^{1/4}
\left(\frac{M_{\star}}{2.4 \ M_{\odot}} \right)^{-5/16}} \\
   {\displaystyle  \left(\frac{r}{1 \AU} \right)^{-1/16}  }  
     \hfill  [\mbox{vis}],  \vspace{0.6cm}\\
 {\displaystyle   0.025
    \left(\frac{M_{\star}}{2.4 \ M_{\odot}} \right)^{-4/7} 
    \left(\frac{L_{\star}}{2.4^4 \ L_{\odot}} \right)^{1/7}  }\\
     {\displaystyle    \left(\frac{r}{1 \AU} \right)^{2/7} } 
       \hfill  [\mbox{irr}],
\end{cases}
\label{eq:aspect}
\end{equation}

The transition radius between these two disk regions is written as
\begin{equation}
    \begin{split}
   r_{\rm tran} =  4.9
    \left( \frac{\dot M_{\rm g}}{10^{-7}  \Msyr} \right)^{28/39}
    \left(\frac{M_{\star}}{2.4 \ M_{\odot}} \right)^{29/39} \\
    \left(\frac{L_{\star}}{2.4^4 \ L_{\odot}}\ \right)^{-16/39}    \AU.
     \label{eq:rtrans}
    \end{split}
\end{equation}

The inner edge of the disk is truncated by the stellar magnetospheric torque \citep{Lin1996OrbitalLocation}.  We approximate the co-rotation cavity radius of the host star 
from \cite{Mulders2015ARates} as:

\begin{equation}
\label{eq:r_in}
    r_{\rm in} {=} \sqrt [3] {\frac{GM_{\star}}{ \Omega_{\star}^{2}}} {\simeq} 0.06 \left( \frac{M_{\star}} {2.4 M_{\odot}} \right)^{1/3} \rm AU,
\end{equation}
where $\Omega_{\star}$ is the stellar spin frequency.

\subsection{Water and silicate snowlines} 
The disk pebbles are assumed to consist of $35 \%$ water-ice and $65\%$ silicates (as in \cite{Liu2019Super-EarthMasses, Johnston2024FormationStars}).  The water snowline is derived by equating the saturation pressure and H$_{2}$O vapour pressure, $P_{\rm sat}(r_{\rm ice}) {=} P_{\rm H_{2}O} (r_{\rm ice})$.  This is because the water snowline is not always at one fixed disk temperature since it also relates to the disk density (Equation \ref{eq:sigma}).  The saturated pressure drops rapidly with disk temperature, and thus the water snowline moves modestly between disk temperatures of $\sim 150{-}190$ K.  This can be approximated to be the disk radii at temperatures of $\sim 170$ K and expressed as: 

  \begin{equation}
\frac{r_{\rm H_{2}O}}{\rm AU} = \begin{cases}
 {\displaystyle   1.56
\left( \frac{\dot M_{\rm g}}{10^{-7} \Msyr} \right)^{4/9}
\left(\frac{M_{\star}}{2.4 \ M_{\odot}} \right)^{1/3}} \\
   {\displaystyle  \left(\frac{\alpha_{\rm g}}{10^{-2}} \right)^{-2/9} \left( \frac{\kappa_{0}}{10^{-2}} \right)^{2/9}  }  
     \hfill  [\mbox{vis}],  \vspace{0.6cm}\\
 {\displaystyle   0.75
    \left(\frac{M_{\star}}{2.4 \ M_{\odot}} \right)^{-1/3} 
    \left(\frac{L_{\star}}{2.4 \ L_{\odot}} \right)^{2/3}  }
       \hfill  [\mbox{irr}]. \\
\end{cases}
\label{eq:r_water}
\end{equation}

The water snowline for this two-component disk model is hence $r_{\rm ice}{=} \rm max (r_{\rm ice, vis}, r_{\rm ice, irr})$. Similarly, the silicate sublimation line from \cite{Johnston2024FormationStars} is adopted as:

  \begin{equation}
\frac{r_{\rm Si}}{\rm AU} = \begin{cases}
 {\displaystyle   0.91
\left( \frac{\dot M_{\rm g}}{10^{-7} \Msyr} \right)^{4/9}
\left(\frac{M_{\star}}{2.4 \ M_{\odot}} \right)^{1/3}} \\
   {\displaystyle  \left(\frac{\alpha_{\rm g}}{10^{-2}} \right)^{-2/9} \left( \frac{\kappa_{0}}{10^{-2}} \right)^{2/9}  }  
     \hfill  [\mbox{vis}],  \vspace{0.6cm}\\
 {\displaystyle   0.06
    \left(\frac{M_{\star}}{2.4 \ M_{\odot}} \right)^{-1/3} 
    \left(\frac{L_{\star}}{2.4 \ L_{\odot}} \right)^{2/3}  }
       \hfill  [\mbox{irr}], \\
\end{cases}
\label{eq:r_sil}
\end{equation}

again, with the silicate sublimation line being the maximum of these two values.  Evaporation of their respective components occurs when pebbles drift inward of their sublimation lines and their pebble mass fluxes decrease accordingly.  As such, the pebbles only lose $\sim 35\%$ of their mass after drifting inside of $r_{\rm H_{2}O}$, so we neglect this size variation within our model.  The released vapour is assumed to be well-mixed with the gas phase of the disk, \textit{e.g.} mass is conserved as it is transferred from the solid (pebble) reservoir to the gaseous reservoir.  We also do not include any enhanced efficiency core accretion processes by icy pebbles near $r_{\rm H_{2}O}$ \citep{Okuzumi2012RapidFormation, Drazkowska2017PlanetesimalLine, Hyodo2021PlanetesimalPebbles}.  That being said, it is evident from Equations \ref{eq:r_water} and \ref{eq:r_sil} that the stellar luminosity plays an important role in setting the locations of these sublimation lines and ultimately dictating the resultant planetary composition.  Lastly, we assume that the dust has already fully grown and adopt a fixed pebble-size regime ($R_{\rm peb}{=}1$ mm) - following the results of \cite{Johnston2024FormationStars} - such that their respective Stokes number is a free parameter.

\subsection{Planet growth and migration}
\label{sec:growth}

We examine the core mass growth of the protoplanetary embryo by pebble accretion. We consider streaming instability as a powerful mechanism that can induce overdensities in the dust, from which planetsimals may form from the collapse of many mm-sized pebbles \citep{Youdin2005STREAMINGDISKS, 
Johansen2007RapidDisks, Johansen2012AddingInstabilities, Bai2010DynamicsFormation, Simon2016THESELF-GRAVITY, Schafer2017InitialInstability, Abod2019TheGradient, Li2019DemographicsInstability}. 
 \cite{Liu2020Pebble-drivenDwarfs} summarized from the literature streaming instability planetesimal formation simulations that the birth masses of the embryos are dependent on their evolutionary time and disk locations, which can be expressed as:

\begin{equation}
    \frac{M_{\rm p0}}{M_{\oplus}} = 6 \times 10^{-2} \left( \gamma \pi \right) ^{1.5}  
    \left (\frac{h_{\rm g}}{0.05} \right ) ^{3} \left ( \frac{M_{\star}}{2.4\ \rm M_{\odot}} \right).
    \label{eq:embryo_mass}
\end{equation}

Consequently, the mass of the embryo has an inherent relation to the stellar luminosity through $h_{\rm g}$, such that $M_{\rm p0} \propto L_{\star}^{3/7}$.  The self-gravity parameter $\gamma$ that quantifies the relative strength between the gravity and tidal shear is given by  
\begin{equation}
    \gamma \equiv \frac{4 \pi G \rho_{\rm g}}{\Omega^{2}_{\rm K}}.
\end{equation}

The solid accretion rate onto the planet's core reads 
\begin{equation}
\label{eq:peb_acc}
    \dot{M}_{\rm PA} = \epsilon_{\rm PA}\dot{M}_{\rm peb} = \epsilon_{\rm PA} \xi \dot{M}_{\rm g},
\end{equation}
where $\epsilon_{\rm PA}$ is the efficiency of pebble accretion, the formulas of which are adopted from \citet{Liu2018CatchingPlanets} and \cite{Ormel2018CatchingPebbles} that include both $2$D and $3$D regimes and expressed as:

\begin{equation}
\label{eq:peb_acc_eff}
    \epsilon_{\rm PA} = \sqrt{\epsilon_{\rm PA, 3D}^{2} + \epsilon_{\rm PA, 2D}^{2}},
\end{equation}
where $\epsilon_{\rm PA, 2D}$ and $\epsilon_{\rm PA, 3D}$ are defined from \cite{Liu2018CatchingPlanets} and \cite{Ormel2018CatchingPebbles}, respectively.  For pebble accretion in the 2D regime from \cite{Liu2018CatchingPlanets}:

\begin{equation}
\begin{split}
    \epsilon_{\rm PA, 2D}{=} \frac{0.32}{\eta} \sqrt{\frac{M_{\rm p}}{M_{\star}} \frac{\Delta \nu}{\nu_{\rm K}}\frac{1}{\tau_{\rm S}}} = 3 \times 10^{-3} \left ( \frac{M_{\rm p}}{0.01 \Me} \right )^{2/3} \\ 
    \times \left ( \frac{\tau_{\rm s}}{10^{-2}} \right )^{-1/3} \left ( \frac{\eta}{3 \times 10^{-3}} \right )^{-1}
\end{split}
\label{eq:pa2d}
\end{equation}

where $M_{\rm p}$ and $M_{\star}$ are the masses of the planet and the star; $\Delta \nu$ is the relative difference between the velocities of the planet and the pebble; $\nu_{\rm K}$ is the Keplerian velocity at the location of the planet; and $\tau_{\rm S}$ is the Stokes number of the pebble.  $\eta = -h_{\rm g}^{2} (\partial \rm ln P / \partial \rm ln r / 2) = (2 - s - q) \ h_{\rm g}^{2}/2$, where $P$ is the gas pressure in the disk.

Concurrently, the pebble accretion efficiency in the 3D regime from \cite{Ormel2018CatchingPebbles} is:

\begin{equation}
    \begin{split}
            \epsilon_{\rm PA, 3D}{=} \frac{0.39}{\eta} \frac{M_{\rm p}}{M_{\star}} = 4 \times 10^{-3} \left ( \frac{M_{\rm p}}{0.01 \Me} \right ) \left ( \frac{M_{\star}}{1 \Ms} \right )^{-1} \\ 
    \times \left ( \frac{h_{\rm peb}}{3 \times 10^{-3}} \right )^{-1/3} \left ( \frac{\eta}{3 \times 10^{-3}} \right )^{-1}.
    \end{split}
    \label{eq:pa3d}
\end{equation}

Whether pebble accretion is in $2$D or $3$D depends on the ratio between the radius of pebble accretion and the vertical layer of pebbles \citep{Morbidelli2015TheCores}.  

The pebble scale height is given by:
 \begin{equation}
 \label{eq:pebh}
    h_{\rm peb} = \sqrt{\frac{\alphat}{\alphat + \tau_{\rm s}}} H_{\rm g},
\end{equation}
where $\tau_{\rm s}$ is the Stokes number of the particles and $\alpha_{\rm t}$ is the turbulent diffusion coefficient, approximately equivalent to the midplane turbulent viscosity when the disk is driven by magnetorotational instability \citep{Johansen2005DUSTTURBULENCE, Zhu2015DustDiffusion, Yang2017ConcentratingInstability}. 

From the pebble core accretion equations defined above arises an important dependence on stellar luminosity.  Namely, the pebble accretion in the 2D regime (Equation \ref{eq:pa2d}) has a $L_{\star}^{-1/7}{-}$dependence through its relation to the scale height of the gas, $h_{\rm g}$.  Similarly, the pebble accretion in the 3D regime (Equation \ref{eq:pa3d}) has a $L_{\star}^{-2/7}{-}$dependence from both the scale height of the gas ($h_{\rm g}$) and the scale height of the pebbles ($h_{\rm peb}$).

A growing planet begins to clear the surrounding gas and opens a partial gaseous annular gap. As a consequence, the inward drift of pebbles stops at the outer edge of the planetary gap and the solid accretion terminates \citep{Lambrechts2014SeparatingAccretion}. Such a planetary mass is referred to as the pebble isolation mass, and we adopt the formula from \cite{Bitsch2018Pebble-isolationGiants} as
 \begin{equation}
   \begin{split}
 M_{\rm iso} = & 16 \left( \frac{h_{\rm g}}{0.03} \right)^3 \left( \frac{M_{\star}}{3 \ M_{\odot}} \right) \left[ 0.34 \left(  \frac{ -3}{ {\rm log_{10}} \alphat}  \right)^4 + 0.66 \right] \\
  & \left[ 1-   \frac{ \partial  {\rm ln } P /  \partial {\rm ln} r +2.5} {6}        \right] M_{\oplus}.
 \label{eq:m_iso}
 \end{split}
 \end{equation}


We only follow the gas accretion when the planet grows beyond $M_{\rm iso}$ (Equation \ref{eq:m_iso}). The gas accretion rate can be expressed as:

\begin{equation}
    \dot{M}_{\rm p, g} = \mathrm{min} \left[ \left( \frac{\mathrm{d}M_{\rm p, g}}{\mathrm{dt}} \right)_{\rm KH} , \left( \frac{\mathrm{d}M_{\rm p,g}}{\mathrm{dt}} \right)_{\rm Hill} , \dot{M}_{\rm g} \right].
\end{equation}

We adopt the gas accretion rate based on \cite{Ikoma2000FORMATIONOPACITY}:

\begin{equation}
\label{eq:kh}
    \left( \frac{\mathrm{d}M_{\rm p, g}}{\mathrm{dt}} \right)_{\rm KH} = 10^{-5} \left( \frac{M_{\rm p}}{10\ \rm M_{\oplus}} \right)^{4} \left( \frac{\kappa_{\rm env}}{1\ \rm cm^{2}\ \rm g^{-1}} \right ) ^{-1} \rm M_{\oplus} yr^{-1}
\end{equation}

where $\kappa_{\rm env}$ is the planet's gas envelope opacity.  We assume $\kappa_{\rm env}{=} 0.05\ \rm cm^{2}\ g^{-1}$ and that it does not vary with metallicity, as in \cite{Liu2019Super-EarthMasses, Johnston2024FormationStars}.

As the planet grows, the accreted gas is further limited by the total amount entering the planetary Hill Sphere, adopted from \cite{Liu2019Super-EarthMasses} below:

\begin{equation}
    \left( \frac{\mathrm{d}M_{\rm p,g}}{\mathrm{dt}} \right)_{\rm Hill} {=} f_{\rm acc} \nu_{\rm H} R_{\rm H} \Sigma_{\rm Hill} = \frac{f_{\rm acc}}{3\pi} \left( \frac{R_{\rm H}}{H_{\rm g}} \right)^{2} \frac{\dot{M}_{\rm g}}{\alpha_{\rm g}} \frac{\Sigma_{\rm gap}}{\Sigma_{\rm g}}. 
\end{equation}

where $\nu_{\rm H}{=} R_{\rm H} \Omega_{\rm k}$ is the Hill velocity; $R_{\rm H}{=}(M_{\rm p} / 3\ \rm M_{\star})^{1/3}$ is the Hill radius of the planet; and $\Sigma_{\rm Hill}$ is the gas surface density near the planet's Hill sphere.  In this paper, we set $f_{\rm acc}{=}0.5$ to be the gas fraction that can be accreted by the planet's Hill sphere.  

We adopt a combined migration rate formula by incorporating both type I and type II regimes \citep{Kanagawa2018RadialPlanet}: 
\begin{equation}
  \dot{r} =  f_{\rm tot} \left( \frac{M_{\rm p}}{M_{\star}} \right) \left( \frac{\Sigma_{\rm g} r^{2}}{M_{\star}} \right) h_{\rm g}^{-2} v_{\rm K},
  \label{eq:mig}
\end{equation}
and the migration coefficient is given by
 \begin{equation}
f_{\rm tot} = f_{\rm I} f_{\rm s} + f_{\rm II} (1- f_{\rm s}) \frac{1} {\left(\frac{M_{\rm p}}{M_{\rm gap}} \right)^2},
\end{equation}
where $M_{\rm p}$ is the mass of the planet, $v_{\rm K}$ is the Keplerian velocity and $f_{\rm I}$ and  $f_{\rm II}$ correspond to the type I and II migration prefactors, respectively. 



\subsection{Stellar luminosity evolution} 
\label{sec:pms_ev}  

\begin{figure}
\centering
\includegraphics[width = \linewidth]
{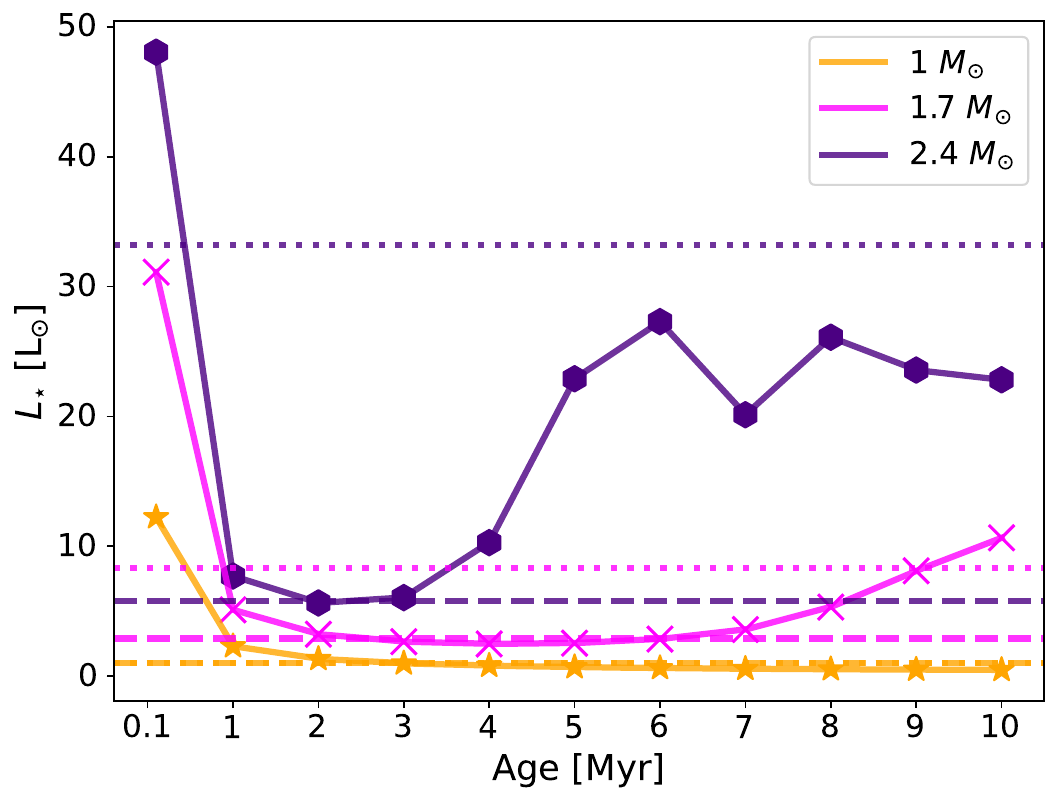}
\caption{Stellar luminosity values for times $0.1{-}10$ Myr, derived from \citep{Siess2000AnStars} for stellar masses $1 \Ms$ (orange), $1.7 \Ms$ (pink), and $2.4 \Ms$ (purple).  The dashed and dotted lines show the fixed luminosity approximations for 1, 1.7, and $2.4 \Ms$ stars ($L_{\star}{=}M_{\star}^{\beta}$) when $\beta{=}2$ and $\beta{=}4$, respectively.}  
\label{fig:lum}
\end{figure}

As a wholly convective PMS star evolves along its Hayashi track until it develops a radiative core, the stellar luminosity evolves in response.  The exact nature of this stellar luminosity evolution has a $M_{\star}{-}$dependence.  Importantly, $L_{\star}$ decreases with age until the main sequence for low mass stars ($M_{\star} \leq 1.5\ \Ms$), but has an upward turn and remains high for more massive stars ($M_{\star} {>} 1.5\ \Ms$), due to their transition to the radiative regime \citep{Palla1993TheStars, Siess2000AnStars, Baraffe2002EvolutionaryAges}.


Pioneering work by Halm and Eddington first demonstrated that the radiative diffusion in stars means that luminosity will depend on the mass to approximately the fourth power: $L_{\star} \propto M_{\star}^{4}$  
 \citep{Kuiper1938THERELATION}.  However, a single exponent is insufficient in explaining the wide range of stellar masses.  \cite{Eker2018InterrelatedRelations} proposed that the stellar mass-luminosity relation for main sequence stars is best expressed by a six piece classical relation:
 
\begin{equation}
     L_{\star} \propto M_{\star}^{\beta}
\label{eq:lum_rel}
\end{equation}
 
with break points for key stellar masses indicative of a shift in the energy transport mechanism (\textit{e.g.} convective to radiative).  $\beta{=}5.7$ for low-mass stars $(0.72 \Ms \leq M_{\star} \leq 1.05 \Ms$) and $\beta{=}4.3$ for intermediate-mass stars $(1.05 \Ms \leq M_{\star} \leq 2.4 \Ms)$.  \cite{Baraffe2002EvolutionaryAges} proposed a $L_{\star} \propto M_{\star}^{1{-}2}$ for modelling low-mass PMS stars.  However, intermediate-mass PMS stars are more luminous than their less massive counterparts.  Thus, we investigate fixed-luminosity prescriptions (Equation \ref{eq:lum_rel}) for our giant planet formation model where $\beta{=}2$ and $\beta{=}4$.   
 
Furthermore, we adopt an evolving stellar luminosity derived from the PMS evolutionary tracks in the Siess models \citep{Siess2000AnStars}.  We select stellar luminosity values at timesteps $0.1{-}10\ \rm Myr$ with solar metallicity ($Z{=}0.01$) for stellar masses in the range of $1{-}2.4 \Ms$ and interpolate the values in between using \textsc{numpy.interp} to return one-dimensional piecewise linear interpolant.  Figure \ref{fig:lum} details how the stellar luminosity evolves over time for 1, 1.7, and 2.4$\Ms$.  These stellar masses were selected as they correspond to important points in the giant planet occurrence rate as a function of $M_{\star}$ (rise, peak, and fall) \citep{Reffert2015PreciseMetallicity, Wolthoff2022PreciseSurveys}, and also so we can contrast our results directly to our previous study which did not include evolving PMS luminosity \citep{Johnston2024FormationStars}.

The $1 \Ms$ stellar luminosity track (orange) consistently decays over time and has the lowest luminosity values overall.  The $1.7 \Ms$ track (pink) initially decreases over time and has only a moderate increase after $5\ \rm Myr$.  Lastly, the $2.4 \Ms$ track (purple) has a sharp decline between $0.1{-}1$ Myr the steepening between $3{-}6$ Myr before flattening out at later times.  The initial sharp decrease followed by a subsequent increase in the stellar luminosity is indicative of a wholly convective star becoming radiative.  This process happens on shorter timescales as stellar mass increases.


\subsection{Midplane temperatures}
\label{sec:temp}


Our planet formation model uses an analytical prescription to describe temperature in the irradiated regime of the disk (Equation \ref{eq:temp}):

 \begin{equation}
 \frac{T_{\rm g}}{ \rm K}  =
  \begin{cases}
 {\displaystyle   500
    \left( \frac{\dot M_{\rm g}}{10^{-7} \Msyr} \right)^{1/2}  \left(\frac{M_{\star}}{2.4 \ M_{\odot}} \right)^{3/8} } \\
   {\displaystyle  \left( \frac{\alpha_{\rm g}}{10^{-2}} \right)^{-1/4} \left( \frac{\kappa_{0}}{10^{-2}} \right)^{-1/4}  \left(\frac{r}{1 \AU} \right)^{-9/8}  }
     \hfill  [\mbox{vis}],  \vspace{0.6cm}\\
 {\displaystyle  236
\left(\frac{M_{\star}}{2.4 \ M_{\odot}} \right)^{-1/7} \left(\frac{L_{\star}}{2.4 \ L_{\odot}} \right)^{2/7}} \\
{\displaystyle   \left(\frac{r}{1 \AU} \right)^{-1/2} \phi_{\rm inc}^{1/4} }
  \hfill  [\mbox{irr, ev}], \vspace{0.6cm} \\
   {\displaystyle  150
\left(\frac{M_{\star}}{2.4 \ M_{\odot}} \right)^{-1/7} \left(\frac{L_{\star}}{2.4 \ L_{\odot}} \right)^{2/7}} \\
{\displaystyle   \left(\frac{r}{1 \AU} \right)^{-3/7} }
  \hfill  [\mbox{irr, fixed}], \\

\end{cases}
\label{eq:temp}
\end{equation}

where $M_{\star}$, $L_{\star}$ are stellar mass and luminosity; $r$ is the radial distance to the central star and $\kappa_{0}$ is disk opacity with a fixed fiducial value of 0.01.  $\phi_{\rm inc}$ is the flaring angle (see \cite{Chiang1997SpectralDisks}), \textit{i.e.} the angle between the incident stellar radiation and the local disk surface and we assume to be fixed at $\phi_{\rm inc}{=}0.05$, which is not too far from realistic values \citep{Dullemond2004TheDisks}.  The viscous temperature $(T_{\rm g} [\rm vis])$ remains the same for both the fixed and evolving luminosity prescriptions.  The irradiated temperature $(T_{\rm g} [\rm irr, ev]$ for the evolving luminosity prescription has been adjusted to fit with the MCMax temperature models described below;  whereas the irradiated temperature equation used for the fixed-luminosity models $(T_{\rm g} [\rm irr, fixed])$ is the same as in \cite{Johnston2024FormationStars}. 
Note that this disk structure is highly simplified given our knowledge of gaps, rings and discontinuities in observed disks - in particular, $\phi_{\rm inc}$ as a positive constant is often untrue \citep{Dullemond2004TheDisks}.  But for the purpose of this paper it is a sufficient assumption.   

In order to ensure this is realistic, we directly compare the evolution of irradiated temperature from our planet formation model against results calculated from radiative transfer models.  We use the Monte Carlo radiative transfer code MCMax \citep{Min2009RadiativeDisks} to iteratively solve for disc heating, cooling, hydrostatic pressure and dust vertical structure (settling) in disks around $1 \Ms$, $1.7 \Ms$ and $2.4 \Ms$ host stars at timesteps 0.1, 1, 3, 5, and 10 Myr, in a physically self-consistent manner.  

An equilibrium between dust settling and vertical mixing is then used to calculate the vertical structure of the dust (see \cite{Dullemond2004TheDisks} for more details).  A self-consistent solution is reached through iterative calculations of thermally supported gas and dust disc structure and the heating of this structure by the central star, which is geometry dependent. We input disc and stellar pre-main sequence evolution parameters for a set of stellar massess and ages into MCMax to obtain snapshots of the 2D disk temperature structure, from which we extract midplane temperature profiles. The aforementioned input parameters are stellar radius and effective temperature of each $M_{\star}$ and age combinations, obtained from the MESA stellar evolutionary tracks \citep{Paxton2011ModulesMESA}.

\begin{figure}
    \centering
    \includegraphics[width=\linewidth]{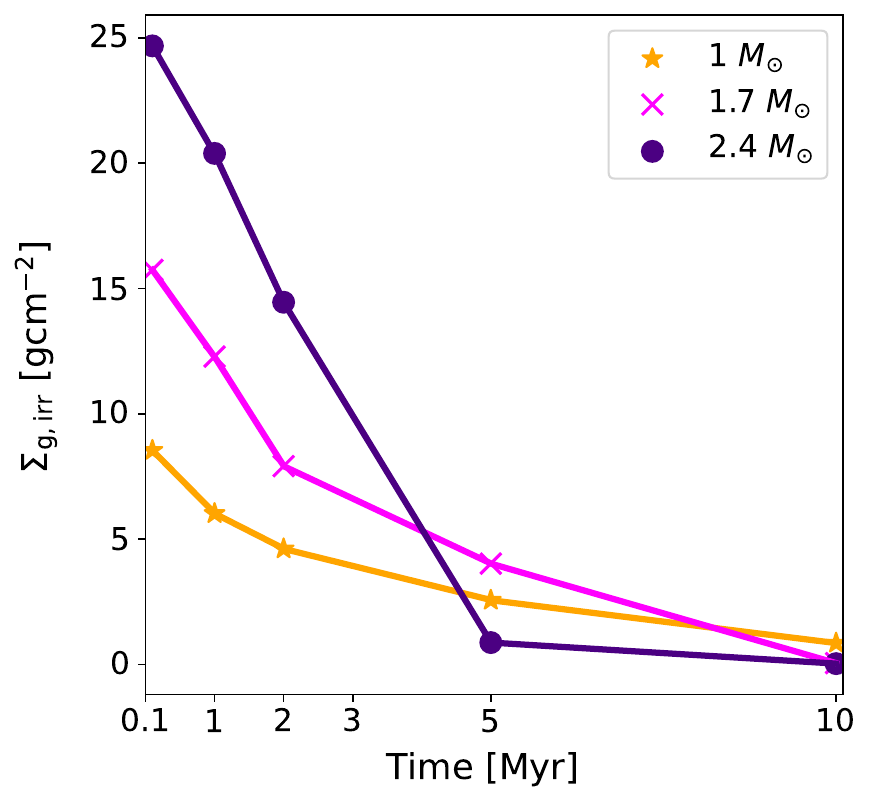}
    \caption{Gas surface density values at $r{=}70$ AU (irradiated regime, Equation \ref{eq:sigma}) in disks around $1 \Ms$ (orange), $1.7 \Ms$ (pink), and $2.4 \Ms$ purple) host stars with evolving PMS stellar luminosity for $0.1{-}10$ Myr  .}
    \label{fig:siggas}
\end{figure}

MCMax takes a given surface density profile (Equation \ref{eq:sigmamcmax} below) and dust-to-gas ratio and assumes hydrostatic equilibrium and $T_{\rm gas}{=}T_{\rm dust}$ to calculate the vertical structure of the disk \citep{Min2009RadiativeDisks}.  The disk surface density profile is adopted from \cite{Andrews2011ResolvedDisks} as:

\begin{equation}
\label{eq:sigmamcmax}
    \Sigma_{\rm gas}{=} \Sigma_{c} \left( \frac{R}{R_{\rm c}} \right) ^{-8} \left[ - \left( \frac{R}{R_{\rm c}} \right) ^{2{-}\gamma} \right], 
\end{equation}

where $R_{\rm c}$ is the critical radius and $\Sigma_{c}$ is the surface density at that critical radius.  To account for disk evolution, we scale the surface density prescription to decrease by certain factors over time according to the surface density evolution in our planet formation models, which involve viscous evolution and photoevaporation. To this end, we extract the surface density of gas at 70~AU in our planet formation model (see Figure \ref{fig:siggas}) at relevant time steps and calculate the fraction by which it decreases, to estimate the factors by which we scale the surface density in the MCMax models. This is important because gas density too affect disc heating significantly, through hydrostatic pressure equation, as shown in \cite{Panic2017EffectsLocation}.

\begin{figure*}
\centering
\includegraphics[width = \linewidth]
{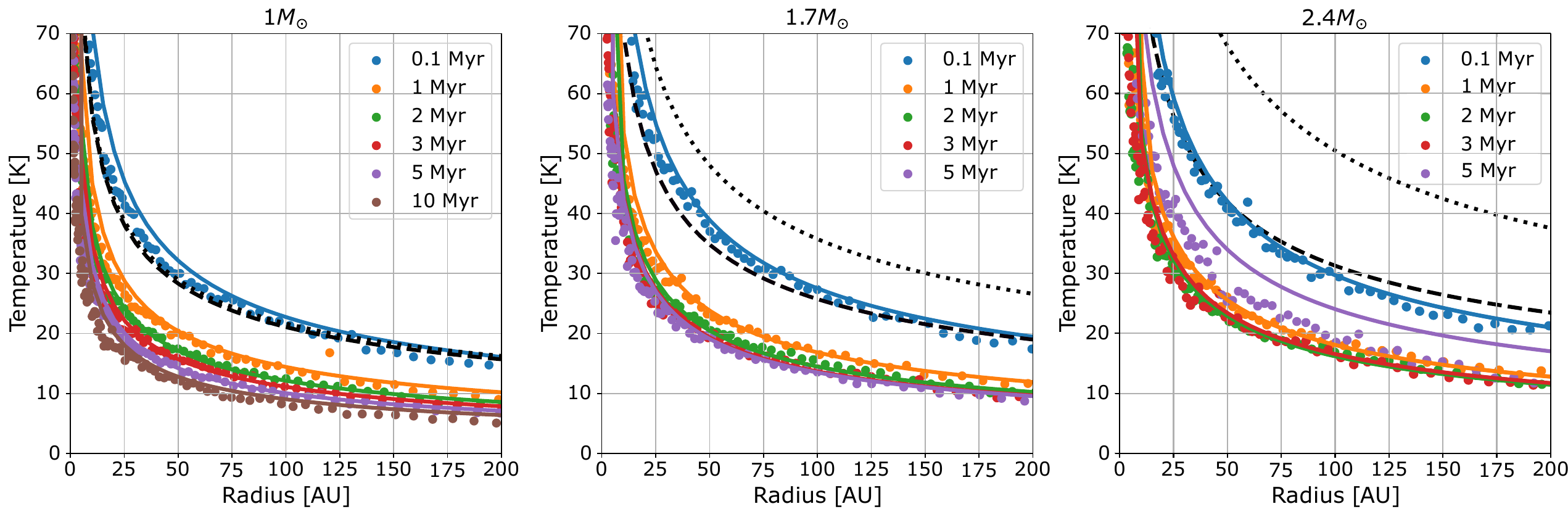}
\caption{Snapshots of the midplane temperature as a function of radius around stars of 1, 1.7, and 2.4 $\Ms$ at different stellar ages.  Each panel represents a different stellar mass.  Each colour represents a different snapshot in time - the scatter data is calculated from the MCMax code and the associated smooth curves are derived from our analytical function for temperature (Equation \ref{eq:temp}).  The black dashed and dotted lines show the temperature of the disk when the assuming a fixed luminosity prescription - \textit{i.e.} $L_{\star} \propto M_{\star}^{\beta}$ - when $\beta{=}2$ and $\beta{=}4$, respectively.  Due to the rate of disk dissipation, we do not include temperature profiles within the disk at times greater than 5 Myr for the $1.7 \Ms$ star the $2.4 \Ms$ star as both disks have mostly dissipated by 8.5 Myr and 5.5 Myr, respectively (see Figure \ref{fig:acc_ev}).}  
\label{fig:temp}
\end{figure*}

From Figure \ref{fig:siggas}, it is clear that the surface density drops at a faster rate in disks around more massive stars.  Additionally, when $t\leq 3$ Myr, the surface density is higher around more massive stars due to disk mass scaling with stellar mass, as described in Section \ref{sec:disk}.  Furthermore from Figure \ref{fig:siggas}, we can see that for the $1.7 \Ms$ host star, the surface density at 70 AU in the disk drops below $0.01 \ \rm gcm^{-2}$ between $5{-}10$ Myr.  Similarly for the $2.4 \Ms$ host star, the surface density drops to just $0.03 \ \rm gcm^{-2}$ by 5 Myr.  Hence, we can conclude that the majority of the disk at large orbital radii has dissipated by 5~Myr. We therefore focus on the first 5~Myr of evolution in this paper, the time-window relevant for giant planet formation. We calculate disc structure up to the time of $\leq 5$ Myr for the $1.7 \Ms$ 
and $2.4 \Ms$ stars. We do not compute 10~Myr temperature profiles because the disk rapidly dissipates between $\sim 5{-}10$ Myr (as described in Figures \ref{fig:acc_ev} and \ref{fig:siggas}).    

 We encompass the percentage decrease in surface density for each disk over time into the MCMax input via the dust mass parameter for each timestep. We assume the gas-to-dust ratio to be 100 in all radiative transfer models.

Figure \ref{fig:temp} shows that our planet formation model analytical function (Equation \ref{eq:temp}, smooth curves) is a suitable approximation for the realistic  midplane temperature profiles obtained from our disc structure modelling using MCMax (scatter data). This is valid for each $M_{\star}$ concerned, as the different panels show, and it can be appreciated that disk temperatures obtained for higher mass stars are comparatively higher, at parity of age. It can also be noted that the temperature steadily decreases over 5~Myr of evolution for  1 and 1.7~$\Ms$ stars but undergoes a distinctive inversion in the case of 2.4 $\Ms$ stars where a rise occurs between 3 and 5~Myr due to the transition to radiative regime of the star as manifested in its luminosity increase shown in Figure \ref{fig:lum}. 

These behaviours were first shown in similar calculations of \cite{Miley2020TheDiscs} who also used MCMax modelling of discs coupled with stellar luminosity evolution on the pre-main sequence. In comparison to their models, which were done for a fixed disc surface density profile for the purpose of isolating the luminosity effects, we use surface density profiles that decrease over time, mimicking disk density evolution implemented in our planet formation models. Lower disc densities (or masses) decrease the hydrostatic support and the disc flattens, decreasing the fraction of starlight captured by the disc \citep{Dullemond2004TheDisks}
, which in turns decreases disc temperature as demonstrated in \cite{Panic2017EffectsLocation}). Therefore keeping consistency in both stellar and disc density evolution when studying disc temperature evolution is important.

We also show the temperatures of disks around 1, 1.7, and 2.4 $\Ms$ stars used in the models without any time evolution of luminosity, i.e., fixed temperature models, exemplifying the $\beta{=}2$ and $\beta{=}4$ versions. Here, $L_{\star} \propto M_{\star}^{\beta}$ at all times in the simulations, resulting in temperature profiles constant over time (seen in black).

While such parametrisation are very roughly similar to the calculations shown for 1.0 $\Ms$ star, for stars of higher masses they largely overestimate disc temperatures. In further sections we will use these prescriptions too, to demonstrate the effect they have on planet formation and to argue against their use in planet formation models, especially where stars more massive than the Sun are considered.  

The $\beta{=}2$ assumption closely resembles the calculated temperature profiles at 0.1 Myr for all stellar masses, but quickly becomes invalid as stars and discs evolve, and planets form in them, on Myr timescales.  The $\beta{=}4$ approximation results in disks far hotter around the $1.7$ and $2.4 \Ms$ stars.  These high disk temperatures have important consequences and challenges for any resultant planet formation that will be discussed in the following sections.  

\begin{table*}
    \begin{tabular}[width=\textwidth]{lclclclclclc|}
        \hline
        \hline
        Disk model & & MCMax code & & Properties \\
        \hline 
        Input parameter & & Input parameter & &  \\
        \hline 
        $L_{\star}$ & Figure \ref{fig:lum} & $R_{\star}$ & MESA stellar evolution track & stellar evolution \\
        & & $T_{\rm eff}$ & MESA stellar evolution track & stellar evolution \\
        $M_{\rm peb}$ & 1 [mm] & a & $0.01 \mu$m ${-}10$ cm & dust size \\
        g/d & 100 & g/d & 100 & gas-to-dust ratio \\
        $M_{\rm disk}$ & $1\%$ of $M_{\star}$ & $M_{\rm dust}$ & ($1\%$ of $M_{\star}$)$\rm (g/d)$ & total disk mass \\
        $R_{\rm d0}$ & 100 [AU] & $R_{\rm c}$ & 100 [AU] & disk size \\
        $\Sigma_{\rm g}(t)$ & Figure \ref{fig:siggas} & $M_{\rm dust}(t)$ & Figure \ref{fig:siggas} & dust mass\\
        $\dot{M}_{\rm g}$ & $8 \times 10^{-8}$ $[M_{\odot} \rm yr^{-1}]$ & & & gas disk evolution\\
        \hline
        \hline
    \end{tabular}
\caption{Disk parameters adopted in our disk model and the  radiative transfer models to calculate temperature profiles in the irradiated disk regime.}
    \label{tab:mcmax}
\end{table*}

Due to the simplistic nature of our irradiated temperature function (Equation \ref{eq:temp}) - which can be expressed as $T_{\rm g}{=}T_{\rm g0} (r/AU)^{p}$ where $p{=}-1/2$ and $T_{\rm g0}$ is the gas temperature at a known distance in the disk - it is insufficient in describing the disk temperature to the accuracy of the radiative transfer models.  \cite{Panic2017EffectsLocation} established that gas mass loss would lead to cooler disk temperatures (alongside less flaring and an increased amount of dust settling in the midplane).  The benefit of Equation \ref{eq:temp} is that it allows for flexibility in investigating a variety of giant planet forming factors, such as characteristic disk size, accretion rates, and birth locations explored in later sections, without the need for full grid of radiative transfer models.

However, this discrepancy is easily addressed as we must scrutinise any formation or significant growth/migration at times later than 8 Myr for $1.7 \Ms$ star and 4 Myr for $2.4 \Ms$ star anyway due to the change in photoevaporation mechanism from X-ray to FUV.  The shift in photoevaporation mechanism corresponds to a convective star becoming radiative.  Higher mass stars have a shorter Kelvin-Helmholtz timescale and therefore develop a radiative core and a hotter photosphere sooner than less massive stars.  \cite{Kunitomo2021PhotoevaporativeStars} found that stars $M_{\star} {<} 1.6 \Ms$ never become hot enough to emit FUV.  We do note that there is a marginal overestimation in the analytical equation versus the MCMax models of around $\sim 5$ K for the 2.4 $\Ms$ at 5 Myr.  Physically, a star of this mass and age would be experiencing this aforementioned shift in photoevaporative mechanism from X-ray to FUV \citep{Kunitomo2021PhotoevaporativeStars}.  Our model is limited to only X-ray photoevaporation so that is likely where this $\sim 5$ K difference between the MCMax model and our model originates.
 e that that the dust and gas are lost at the same rate (by our aforementioned process of evolving the surface density in our models and extrapolating that to the gas loss).

\begin{figure}
    \includegraphics[width = \linewidth]
    {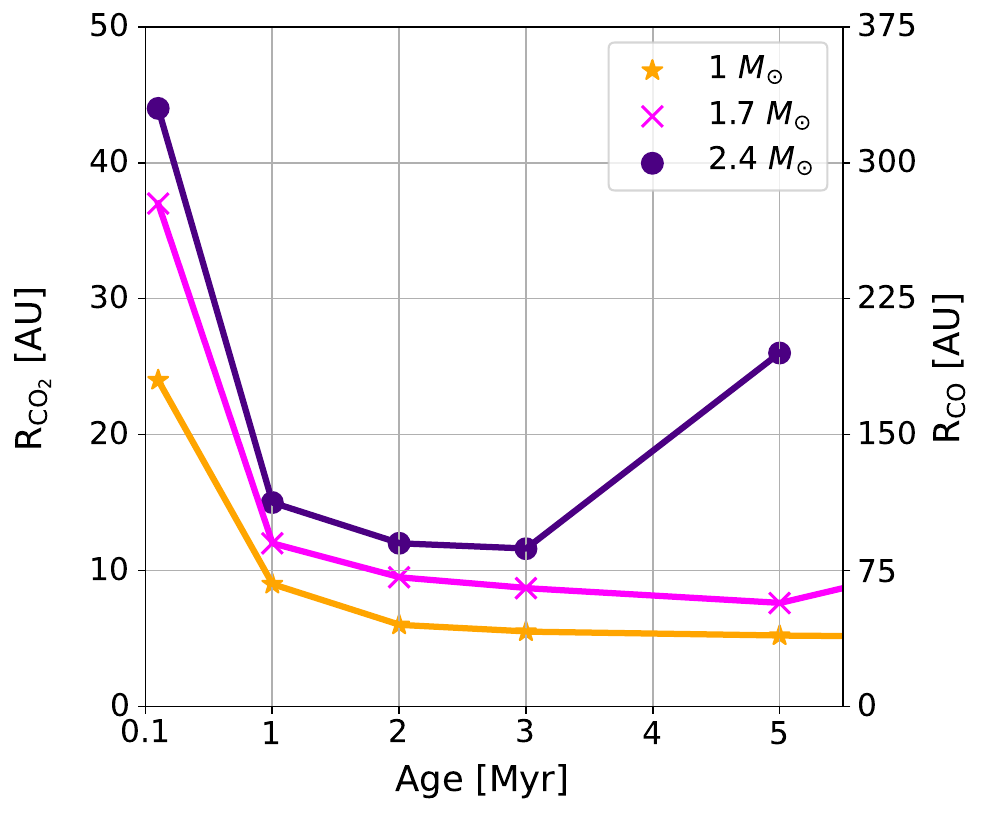}
    \caption{Position of the CO$_{2}$ ($\rm T{=}47 \ \rm K$, left axis) and CO ($\rm T{=}20\ \rm K$, right axis) midplane snowline locations in disks around stars of $1 \Ms$, $1.7 \Ms$, and $24 \Ms$, for $t{=}0.1$, 1, 2, 3, and 5 Myr. Each model has a total disk mass of $0.1 M_{\star}$.}
    \label{fig:cotrack}
\end{figure}

As in \cite{Miley2020TheDiscs}, a direct consequence of the variation in midplane temperature is the change in the location of snowlines within the disk.  Realistically, disk temperature at various snowlines are not universal for each species but depends slightly on local pressure and concentrations \citep{Murillo2022ModelingCores}.  Laboratory experiments have found that CO desorption energies can vary by $\geq 50\%$ depending on water coverage and ice structure when absorbed \citep{Noble2012ThermalCoverages}.  We do not explore or track the dust grains in this paper.  We also do not consider "enhanced" pebble accretion near the water snowline due to icy "sticky" pebbles.  Instead, we adopt the aforementioned static $T_{\rm CO}$ and $T_{\rm CO_{2}}$ with the goal to examine global trends across $M_{\star}$ and stellar age.  However, we do include a dynamical water and silicate sublimation lines, calculated by saturation pressure and detailed in Section \ref{sec:disk}.

Figure \ref{fig:cotrack} shows the CO and CO$_{2}$ midplane snowline locations, assumed to be $\rm T{=}20$ K and $\rm T{=}47$ K, respectively, around host stars of different masses.  While we are not attempting to replicate any specific observed disks, especially given their many known structural complexities and specific CO freeze out temperatures (\textit{e.g.} 17 K for TW Hya; 25 K for HD 163296 \citep{Qi2015CHEMICALDISK}).  Instead, Figure \ref{fig:cotrack} illustrates our model estimations for the CO and CO$_{2}$ snowline locations and how this might compare to observations around similar stellar masses. 

As established in \cite{Miley2020TheDiscs} and discussed in our Section \ref{sec:temp}, disks around less massive stars cool at a faster rate than around more massive stars due to their stellar luminosity consistently decreasing with age during the PMS phase.  In Figure \ref{fig:cotrack} we can see that within 1 Myr, the CO snowline around the $1 \Ms$ (orange) is within 50 AU of the host star.  Conversely, intermediate-mass stars increase in stellar luminosity as they become radiative during the PMS phase, the CO snowlines around the $1.7 \Ms$ and $2.4 \Ms$ are pushed to greater radial distances at $t > 5$ Myr and $t > 3$ Myr, respectively.  This has important composition consequences on giant planet formation, specifically where and when successful giants accrete the majority of their gaseous material.  

Similarly, the CO$_{2}$ snowline in the disk around a $1 \Ms$ host star quickly shifts inward in response to the cooling disk, reaching $\sim 5$ AU in just 3 Myr.  This is contrasted in the disks around the more massive host stars - $1.7 \Ms$ and $2.4 \Ms$ - where at times later than 5 Myr and 3 Myr respectively, the CO$_{2}$ snowlines once again shift outward in response to the warming of the disk due to the increase in stellar luminosity.  By 5 Myr, the $R_{\rm CO2}$ snowline is found at distances around 5 times greater in the disk around a $2.4 \Ms$ star than around a $1 \Ms$.   

This is in-keeping with trends reported in current literature.  It has been shown that disks around T Tauri stars have much weaker CO abundances than anticipated, which may be due to carbon- and oxygen-rich species being bound in ice.  Chemical conversion into more complex ices \citep{Bosman2018COSurface, Agundez2018TheStars} and radial transport \citep{Krijt2018TransportDrift} could both explain the low CO fluxes detected in T Tauri disks.  Meanwhile, since Herbig disks are warmer due to their more luminous host star \citep{Miley2020TheDiscs}, less CO conversion occurs as a result \citep{Bosman2018COSurface}.

\section{Results}
\label{sec:gromap}

\label{sec:mdotg}

\subsection{Setup}
\label{sec:pgro}

\begin{table}
    \centering
    \begin{tabular}{lclclclclclc|}
        \hline
        \hline
        Parameter   &   Description  &   \\ 
            \hline
        disk model   &  viscously heated + stellar irradiation \\
        $\rm M_{\star}$  & $1$, $1.7$, \& $2.4$ $[\Ms]$ \\ 
        $L_{\star}$ & $\propto M_{\star}^{2}$ and evolving (Figure \ref{fig:lum}) \\
        $\rm Z_{\star}$ & $0.0$ [Fe/H] \\
         $\dot{M}_{\rm g0}$ &  $10^{-7}$ $[\rm M_{\odot} \rm yr^{-1}]$ \\
         $ R_{\rm d0}$ &  160 [AU] \\
          $ r_0$  &  $1{-}150$ [AU] \\
           $ t_0$  & $0.5\ \rm [Myr]$ \\ 
            $R_{\rm peb}$ &  $1 \rm \ [mm]$ \\
            $ M_{\rm p0}$ &  \eq{embryo_mass} $[\Me]$\\
      \hline
        \hline
    \end{tabular}
\caption{Adopted parameter distributions for the embryo growth tracks in Figure \ref{fig:pmig_new}. }
    \label{tab:pmig}
\end{table}

In this section, we focus on identifying the regions that directly contribute the bulk of the material used to build the giant planets in our models, \textit{i.e.} the locations where runaway gas accretion occurs. As exemplified by Figure \ref{fig:cotrack} and demonstrated in the work of \citet{Miley2020TheDiscs}, the migration of the snowlines of important C- and O- carrier species can yield different C/O rations of accreted material depending on time and location of the runaway mass growth onto the planet. This could have important implications for the C/O ratios observed in giant planets, especially should our work show clearly distinct scenarios. 
We investigate the times and locations where the successful giant planets accumulate the bulk of their mass and examine this in the context of available chemical abundances at that stage of disk evolution and location within the disk (see Figure \ref{fig:cotrack}). 

In order to identify the times and location of runaway growth we explore the individual growth and migration of single embryos with initial accretion rate $\dot{M}_{\rm g0}{=}10^{-7} \Ms \rm yr^{-1}$ and initial disk size $R_{\rm d0}{=}160$ AU around 1, 1.7, and $2.4 \Ms$ host stars, but focusing only on the early birth time $t_{0}{=}0.5$~Myr.  The aim is to examine the impact of PMS stellar luminosity alone on the successful birth locations of giant planets and examine the ramifications this may have on planetary composition.

We assume all pebbles to have a constant size of 1 mm (supported by mm-wavelength observations of disks \citep{Draine2006ONDISKS, Perez2015GRAIN25}).  We focus on this pebble size as it is the most efficient for pebble accretion \citep{Johnston2024FormationStars}.  In this regime, pebbles are all 1 mm but have a varying Stokes number depending on their location in the disk and the stage of disk evolution (Equations \ref{eq:peb_acc}- \ref{eq:pebh}).  At later stages and greater disk radii, the pebbles have larger Stokes number which results in them essentially becoming de-coupled from the movement of the gas and behaving like planetesimals.  Consequently, less efficient pebble accretion occurs \citep{Ormel2010TheDisks, Ormel2018CatchingPebbles}.  The initial mass of the embryo is calculated using Equation \ref{eq:embryo_mass}, via streaming instability.

\subsection{Growth and migration of embryos born at 0.5 Myr }
\label{sec:pgro_emb}

\begin{figure*}
    \includegraphics[width=\textwidth]{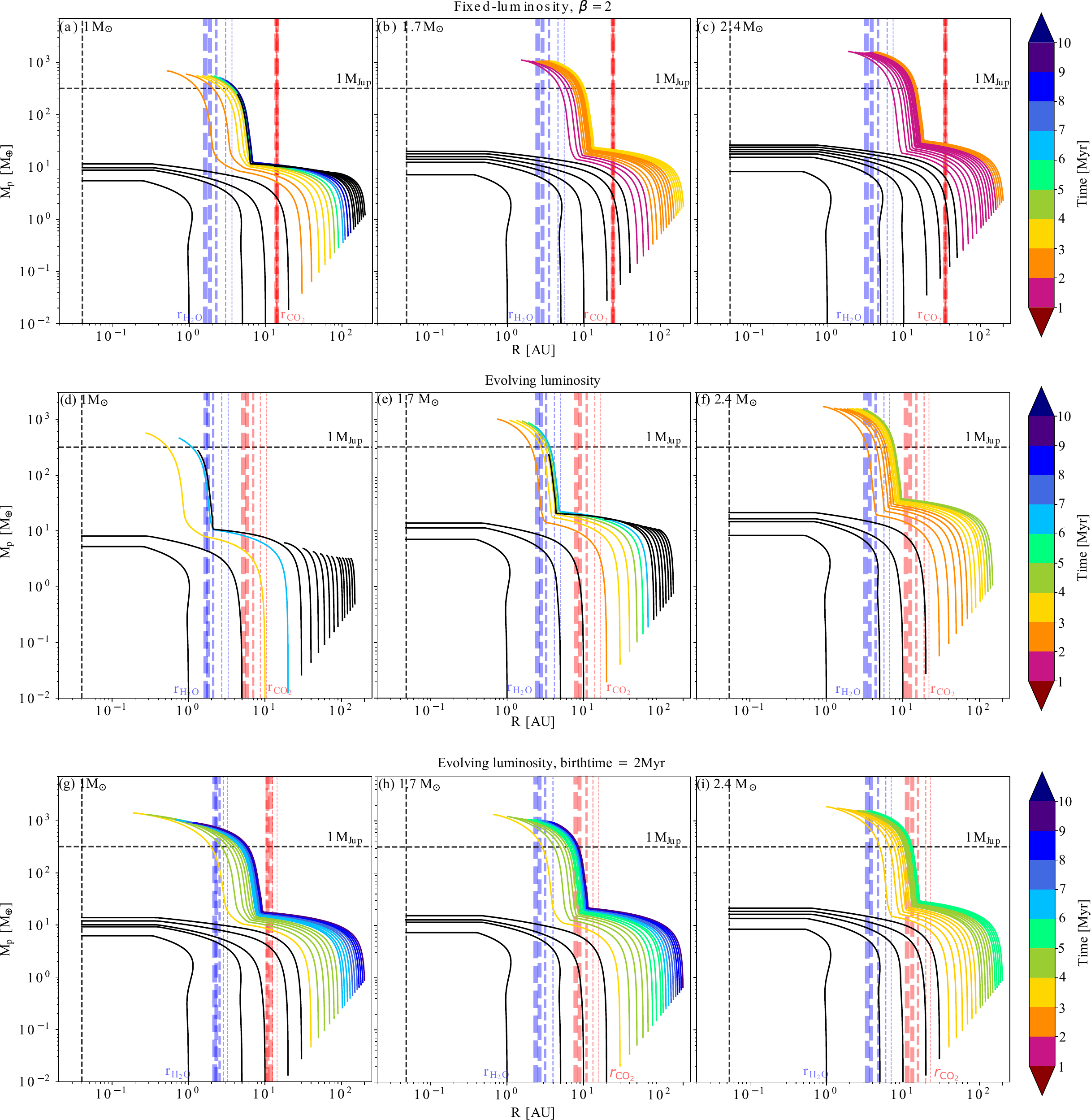}
    \caption{Migration and growth tracks of planets as functions of planet location $r_{p}$ and planet mass $M_{\rm p}$ around stars of (a, d, g) $1\ \rm M_{\odot}$, (b, e , h) $1.7\ \rm M_{\odot}$, and (c, f, i) $2.4\ \rm M_{\odot}$.  The top row is the fixed-luminosity case for $L_{\star} \propto M_{\star}^{\beta}$ when $\beta{=}2$; the middle row is our evolving luminosity prescription; and the bottom row is our evolving luminosity prescription for embryos born at 2 Myr.  The initial accretion rate is $\dot{M}_{\rm g0}{=}10^{-7} \Ms \rm yr^{-1}$.  The initial characteristic disk sizes are all $R_{\rm d0}{=}160$ AU.  The horizontal dashed line is located at $318 \Me$ or $1 \rm M_{\rm Jup}$ to denote our threshold minimum mass for a giant planet. The black vertical dashed line is the inner edge of our disk, the location where inward migration and planetary mass growth ceases (Equation \ref{eq:r_in}).  The blue and red dashed lines are the locations of the $\rm H_{2}O$ and $\rm CO_{2}$ snowlines for times 0.5, 1, 2, 3 Myr, the thickness of the line increasing with time accordingly.  The black tracks mean the planet does not surpass $1 \rm M_{\rm Jup}$.  The successful giant planets are coloured with respective to the time when they surpass $300 \Me$, and can be found in the colour bar on the right hand side.}
    \label{fig:pmig_new}
\end{figure*}

Figure \ref{fig:pmig_new} compares the growth of individual embryos in the fixed luminosity scenario (top row) to that in the evolving luminosity scenario (middle and bottom row), for three different stellar masses. The different tracks show the growth of the planetary mass as a function of orbital distance, for a range birth locations ($r_{0}{=}1{-}150$ AU). We particularly focus on the embryos which successfully grow over the mass threshold to qualify as a giant planet, which we colour code according to the time when they accumulate the bulk of their mass. Besides of interest in the time and place when the bulk of the mass is accreted, called \textit{growth location} hereafter, we are also interested in how this growth location relates to the locations of the snowlines of the major O- and C- carrier species such as H$_2$O, CO$_2$ and CO at the corresponding times, as this informs us of the C/O ratio of the material accreted which we indicate in Figure \ref{fig:pmig_new}.   

In the top two rows of Figure \ref{fig:pmig_new} (a-f), we are focusing on the fate of embryos born at 0.5~Myr, seen in the maps as the bottom boundary in each panel. Each panel has a consecutive group of successful embryos shown in colour, of which the ones born closer in always become giant planets sooner and terminate at closer-in orbits than any that are born further out. The giants in the top panels form at comparatively earlier times than in the corresponding evolving luminosity scenarios shown in the bottom panels. There is some variety in the span of birth locations of the successful embryos, and consequently in the span of their growth locations, but as a rule of thumb we find most of the growth occurring between the H$_2$O and CO$_2$ snowlines, where C/O is ${<} 1$ \citep{Oberg2011TheAtmospheres}.

In the case of fixed luminosity $\beta{=}2$ (top row, Figure \ref{fig:pmig_new}) giant planets accumulate the bulk of their mass in just $1{-}3$ Myr. 
Successful embryos are born as far as the outermost orbital distance explored here - 160~au in the case of the more massive stars, and they migrate and assemble very fast, with growth zone typically around 20~au. This mass accumulation occurs inside the $r_{\rm CO_{2}}$ snowline but outside the water ice line $(r_{\rm H_{2} O})$.   For 2.4$\Ms$ star an embryo born at 100~au migrates to 20~au and grows to 1000 M$_{\oplus}$ in under 2~Myr, for example.

In the case of the evolving luminosity (middle row, Figure \ref{fig:pmig_new}), it is overall harder to form giant planets than in the fixed luminosity case - it takes a longer time, typically more than 3~Myr. This is mainly because the growth zones are much closer to the star, and even though the birth distances of the successful embryos are closer too, these embryos need to migrate for a longer time to reach them.  Giant planets are still efficiently formed, to masses of 1000 M$_{\oplus}$ around more massive stars where the embryos come from large orbital distances, but this time they have to travel all the way to several au from the star before they can undergo runaway growth. This is discussed in more detail in Section \ref{sec:comp}.

\subsection{Growth and migration of embryos born at 2 Myr} 
\label{sec:2myr}

In the bottom row of Figure \ref{fig:pmig_new}, we are focusing on the fate of embryos born at 2 Myr.  Each panel has a consecutive group of successful embryos, of which the ones born closer in always become giant planets sooner and terminate at close-in orbits than any that are born further out.  There is some span of both birth and growth locations of the successful embryos. but most growth occurs between the H$_{2}$O and CO$_{2}$ snowlines.

Most giant planets now accumulate the bulk of their mass in $3{-}5$ Myr (or 1-2 Myr from the 2 Myr embryo birth time).  Successful embryos are born at large orbital radii around the more massive stars, and migrate and assemble very fast, with growth zones typically around 20 AU.  The mass accumulation occurs inside the CO$_{2}$ snowline but outside the water ice line. The growth and migration tracks in the bottom row of Figure \ref{fig:pmig_new} now much more closely resembles the fixed luminosity $\beta{=}2$ case in Figure \ref{fig:pmig_new} (top row).  Overall, giant planet formation is clearly far easier for embryos born at 2 Myr than at 0.5 Myr in the evolving luminosity case (Figure \ref{fig:pmig_new}, middle row).  This follows our reasoning discussed in Section \ref{sec:comp}.  By 2 Myr, the star is less luminous than at earlier birth times, meaning that the scale height of the disk is more compressed (Figure \ref{fig:lumdiag}), allowing for increased pebble accretion efficiency (Equation \ref{eq:peb_acc_eff}) and thus faster solid core assembly.

The timescale required to build the solid core mass of a giant planet is often the biggest evolutionary barrier.  A balance needs to be struck between said solid core having time to assemble enough mass in order to undergo runaway gas accretion and there still being enough gas present in the disk to do so.  However, this is circumvented by the fact that we know that giant planet formation is already well underway in Class II discs ($t \geq$ 1 Myr) \citep{Andrews2018TheOverview, Manara2018WhyPopulation, Benisty2021APDS70c, Izquierdo2022ADiscminer, Testi2022TheYSOs}.  While giant planet formation is easier for embryos with 2 Myr birth time, it is likely to still be incredibly challenging physically.  In the case of the 2.4 M$_{\odot}$ star, the quickest giant assembly time is still 2 Myr, putting the age of the disk at 4 Myr.  At this age, the stellar luminosity of this intermediate-mass star is already beginning to increase as it becomes radiative (Figure \ref{fig:lum}), meaning that FUV would replace X-ray as the dominant mass-loss mechanism.  Since FUV photoevaporation drains small grains with the gas which could impede the assembly of the solid core \citep{Nakatani2021PhotoevaporationRemnants}.

\section{The impact of stellar luminosity}
\label{sec:comp}

\begin{figure*}
    \centering
    \includegraphics[width=\linewidth]{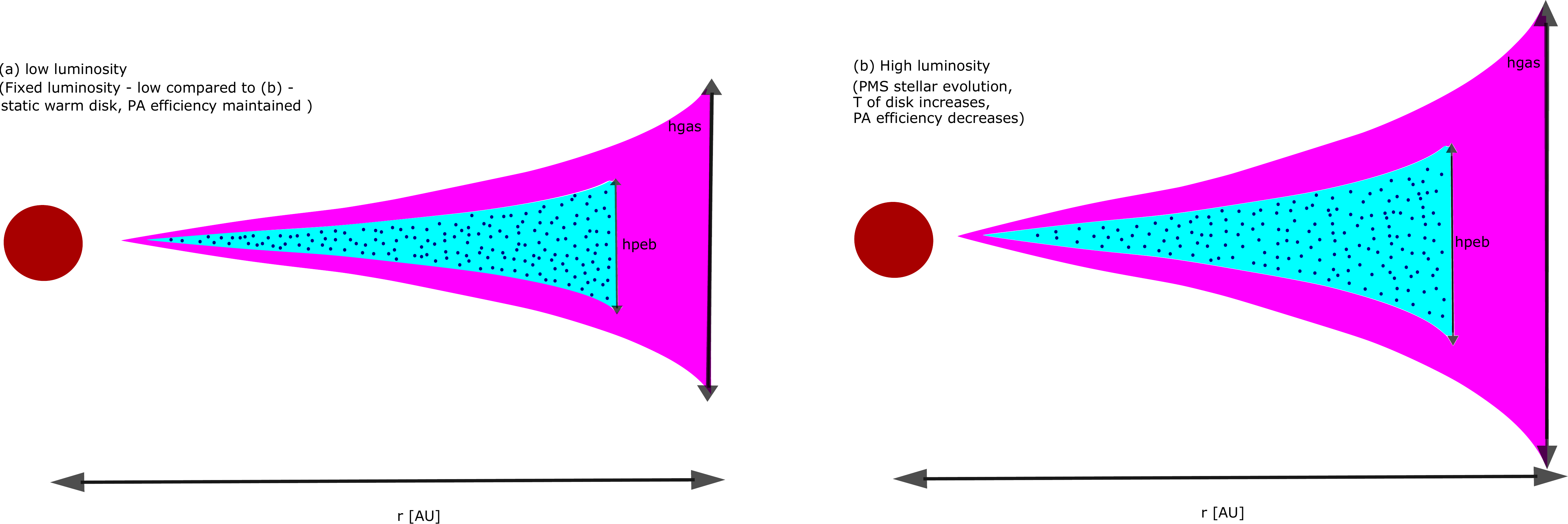}
    \caption{Cartoon of the physical disk and formation processes directly affected from adopting either (a) the fixed luminosity where $\beta{=}2$ defined in Figure \ref{fig:lum} or (b) the evolving PMS stellar luminosity.  If the stellar luminosity is high, the scale height of the gas disk and pebbles increases causing the pebble accretion efficiency to decrease.  Conversely, if the stellar luminosity is low (and kept constant), the scale height of the gas and pebbles remain relatively small which increases pebble accretion efficiency.  This means that fast solid core assembly of planets is favoured in systems with less luminous host stars.   }
    \label{fig:lumdiag}
\end{figure*}

We find that the inclusion of time-dependent PMS stellar evolution significantly impacts the birth location of embryos that grow into successful giant planets.  
This has a compounding affect on where and when in the disk these giant planets undergo runaway gas accretion to accumulate the bulk of their gas mass, which in turn, may impact the overall chemical composition of the giant planets.


From Figure \ref{fig:pmig_new}, it is clear that changing the stellar luminosity prescription - and consequently, the disk temperature - has a profound impact on the locations and times where successful giant planets are born and accrete the bulk of their mass.  This is because the stellar luminosity, $L_{\star}$, is an integral component that governs planetary growth and disk evolution.  Scale height of the gas disk - $h_{\rm g}$ - is one aspect with an intrinsic relation to $L_{\star}$.

From Equation \ref{eq:aspect}, in the regions of the disk heated by irradiation from the host star, $h_{\rm g}$ has a $L_{\star}^{1/7}{-}$dependence.  This means that the scale height of the gas increases with the stellar luminosity to the power of $1/7$.  Since $h_{\rm g}$ - and $h_{\rm peb}$, scale height of the pebbles - is involved in the rate of pebble accretion onto the planet's core (Equations \ref{eq:peb_acc}{-}\ref{eq:pebh}), $L_{\star}$ plays a role in setting the time for a solid planet core to assemble.  Simply, if $L_{\star}$ is large, both $h_{\rm g}$ and $h_{\rm peb}$ are also large, which decreases the solid core accretion rate onto the planetary core.

Furthermore, $h_{\rm g}$ - and so, $L_{\star}$ - are involved in setting the pebble isolation mass $(M_{\rm iso})$ in a system.  This is another property that increases with increasing $h_{\rm g}$, and so has a $L_{\star}$-dependence.  To summarise, it becomes challenging for giant planets to form around very luminous stars because the rate of solid accretion onto the planet via core accretion is decreased and they must accumulate more solid material to undergo runaway gas accretion than environments around less luminous host stars.   A cartoon of this physicality is shown in Figure \ref{fig:lumdiag} (not to scale).  

Another aspect of planet formation intertwined with the $L_{\star}$ through $h_{\rm g}$ is the rate of migration, $\dot{r}$ (Equation \ref{eq:mig}).  If $L_{\star}$ is large, the rate of migration decreases, or the migration timescale increases \citep{Johnston2024FormationStars}:  

\begin{equation}
\label{eq:tmig}
    \tau_{\rm mig}{=}f_{\rm tot}^{-1} \left ( \frac{M_{\star}}{M_{\rm p}} \right) \left( \frac{M_{\star}}{\Sigma_{\rm g} r^{2}} \right) h_{\rm g}^{2} \Omega_{\rm K}^{-1} \propto M_{\star}.
\end{equation}

Meanwhile, $L_{\star} \propto M_{\star}^{2}$ might first appear like it shares many similarities with the evolving luminosity prescription.  In Figure \ref{fig:lum}, the $L_{\star} \propto M_{\star}^{2}$ and evolving luminosity values share common points for (i) $1{-}10$ Myr for the $1 \Ms$ star (ii) $2{-}7$ Myr for the $1.7 \Ms$ and (iii) $2{-}3$ Myr for the $2.4 \Ms$ star.  Furthermore, in Figure \ref{fig:temp}, the $\beta{=}2$ luminosity prescription provides a disk temperature that is in good agreement with the MCMax calculations for disks around 1, 1.7, and $2.4 \Ms$ stars at 0.1 Myr (a warm disk approximation).  
However, Figure \ref{fig:pmig_new} reveals very different growth rates for giant planet formation. 

Since the embryos of our successful giant planets in our planet formation model are now born at greater radial distances within the disk, this changes some of our core composition assumptions.  Namely that the majority of embryos would now be formed beyond both the water snowline and some the CO$_{2}$ snowline, even around more massive stars with warmer disks.  The overall warmer temperature of disks around more massive stars would be reflective in a larger reservoir of water and simple organics in the inner regions combined with a lower abundance of ices in the outer disk, as seen in infrared and (sub)-millimetre observations carried out on disks around Herbig Ae/Be stars in \cite{Agundez2018TheStars}.     
However, in terms of overall giant planet composition, the solid core only accounts for a relatively small fraction of the overall composition.  As a whole, giant planets are mostly comprised of hydrogen and helium.  The ratio of carbon to oxygen (C/O) of an exoplanet atmosphere is often used as the link to an exoplanet's formation site due to the radially gradient nature of C/O ratios in the disk midplane in both gas and ice \citep{Oberg2010TheData}.  

The snowlines of $r_{\rm CO_{2}}$ are fixed in the fixed luminosity case (constant temperature). These snowlines are located in the irradiation-heated outer region of the disk and in the case of the evolving stellar luminosity, they migrate inwards as the disc cools. But, across our models shown in Figure \ref{fig:pmig_new} giant planet growth zones are always inside the  $r_{\rm CO_{2}}$ snowlines. However, more relevant is their location with respect to the $r_{\rm H_{2}O}$ snowline. In both luminosity scenarios explored, the snowlines of $r_{\rm H_{2}O}$ move inwards over time as this region of the disc is heated viscously, so there is no significant difference between the fixed and evolving luminosity cases. What is different, though, is that in the evolving luminosity case the growth zones are so much closer to the star than in the fixed luminosity case, that in some cases we can see formation inside the $r_{\rm H_{2}O}$  snowline as exemplified in the case of $1.0 \Ms$ star models. Such planets are likely to have a low C/O ratio, as they are accreting O-rich gas where all the key O-bearing species are in the gas phase. 

We have focused on the composition of the gas accreted during runaway growth as it directly influences the observable atmospheric C/O ratio.  However, we note that for the bulk C/O ratio of a Jupiter-mass planet, the solid accretion during the core-building phase often dominates the budget \citep{Mordasini2016TheJupiters}.  The composition of these solids, determined by the birth location and migration path of the embryo across various ice lines, is therefore equally important.  A complete interpretation of planetary composition would require tracking both solid and gas accretion histories with respect to the icelines, which is beyond the scope of this study.

Giant planets form incredibly quickly in the case of $\beta{=}2$ stellar luminosity.  When the stellar luminosity evolves, the giant planets form at a more sedate pace due to the many environmental challenges described above.  This does highlight the vital need to incorporate evolving PMS stellar luminosity into planet formation models as it directly impacts many physicalities of the disk that are key to the underlying processes governing giant planet formation and their resulting planetary composition.

The other overall trends concluded from our previous study in \cite{Johnston2024FormationStars} that did not include an evolving PMS stellar luminosity component, nor the matching of the analytical equation for midplane temperature with radiative transfer models, remain the same.   Namely, giant planets prefer to form in systems with large initial disk radius $(R_{\rm d0}{=} 160$ AU) and high initial accretion rate ($\dot{M}_{\rm g0}{=}10^{-7} \Ms \rm yr^{-1}$) - and thus large disk mass due to $M_{\rm d} \propto \dot{M}_{\rm g}$ correlation.

The main goal of this study was to isolate the impact of stellar luminosity on giant planet formation, with a focus on how this varies between stellar masses.  However, that is not to say that stellar luminosity dominates over other planet forming parameters, such as accretion rate and viscosity.  This is because, as discussed above, the impact of changing the stellar luminosity to an evolving prescription versus a fixed assumption has a relatively minor affect linked to the scale height of the pebbles, \textit{e.g.} $h_{\textrm{peb}} \propto L_{\star}^{1/7}$, which results in less efficient pebble accretion during the solid core building phase.

For example, the solid core accretion rate onto the planet's core (Equation \ref{eq:peb_acc}) also depends on the accretion rate ($\dot{M}_{\textrm{g}}$.  Ergo, accretion rate is likely to overcome any decreased efficiency caused by the high luminosity state.  Nevertheless, our evolving luminosity prescription allows for improved modelling of the temperature variations within the irradiated-dominated regions of the disk compared to the fixed stellar luminosity assumption.

Our model assumes that the disk is in a quasi-steady state at each timestep (see Section \ref{sec:method}.  While the stellar luminosity of a $2.4 \textrm{M}_{\odot}$ star changes significantly over $sim$ Myr, the viscous timescale for $r \lesssim$ 10 AU is much shorter ($\sim 10^{4}{-}10^{5}$ years).

Our model utilizes the non-rotating, non-accreting stellar evolution of \cite{Siess2000AnStars}.  We acknowledge that stellar rotation and accretion may impact the structural evolution and contraction of a PMS star \citep{Eggenberger2019RotationInteriors}, potentially altering the PMS luminosity evolution track.  Incorporating these affects, especially for more massive stars, is an important area for future work.

Furthermore, our model does not include accretion luminosity, \textit{e.g.} $L_{\textrm{acc}} = G M_{\star} / 2R_{\star}$.  For our typical accretion rates ($\dot{M}_{\textrm{g}} \sim 10^{-7} \textrm{M}_{\odot}$ yr${-1}$, $L_{\textrm{acc}}$ is comparable to or can exceed the stellar photospheric luminosity, $L_{\star}$ in the  early disk phase.  This additional heating source could further increase disk temperatures, potentially pushing ice lines to greater radial extents and impacting the pebble accretion efficiency.  While not included in this work, it could alter the stellar evolution itself \citep{Baraffe2017Self-consistentDwarfs}, adding another layer of complexity for future models to consider.

\section{Conclusions}
\label{sec:conc}
In this work, we employ and extend the pebble-driven core accretion models \citep{Liu2019Super-EarthMasses, Johnston2024FormationStars} to study how giant planets form around realistic, evolving stars during the pre-main sequence phase.  To this end, incorporate PMS luminosity evolution and focus on the stellar mass range of $1{-}2.4 \Ms$ (Figure \ref{fig:lum}).

We find that employing an evolving PMS stellar luminosity prescription is instrumental in setting many physical qualities of the disk that govern the rate at which planets can form.  Importantly, the scale height of the gas is larger when the disc is warmer, \textit{i.e.}, when the star is more luminous.  This goes on to make the scale height of the pebbles larger, and the rate of core accretion onto the planetary core slower (Figure \ref{fig:lumdiag}).  Additionally, when $L_{\star}$ is higher and the scale height of the disk is large, the planet must accumulate more solid material to surpass the pebble isolation mass in order to undergo runaway gas accretion than when $L_{\star}$ is lower.  The rate of planetary migration is also slower around more luminous stars.  Simply, the rate of planet formation is slower around more luminous stars than around less luminous stars (Figure \ref{fig:pmig_new} and \ref{fig:lumdiag}).  

Lastly, we highlight the importance that PMS stellar luminosity plays in setting the location of volatile freezeout.  We calibrate our disk temperature such that it corresponds to the MCMax radiative transfer calculations for our initial conditions (Figure \ref{fig:temp}).  From this, we can approximate the freezeout locations for CO and CO$_{2}$ (Figure \ref{fig:cotrack}).  Consequently, we can examine the locations and times within the disk where our model giant planets accumulate the bulk of their mass (Figure \ref{fig:pmig_new}).  We find that lower, solar mass stars have giant planet growth located inside the CO$_{2}$ snowline.  This result supports observational findings that giant planets around more massive stars tend to be oxygen-poor but metal rich \citep{Madhusudhan2014H2OJupiters, Welbanks2019K} (Figure \ref{fig:pmig_new}).


\section*{Acknowledgements}

HFJ was supported by a Royal Society Grant (RF-ERE-221025) and the Science and Technology Facilities Council (Project Reference: 2441776). OP acknowledges support from the Science and
Technology Facilities Council, grant number ST/T000287/1.  BL is supported by National Key R\&D Program of China (2024YFA1611803), National Natural Science Foundation of China (Nos. 12222303 and 12173035), the start-up grant of the Bairen program from Zhejiang University.  This research was supported by the Munich Institute for Astro-, Particle and BioPhysics (MIAPbP), which is funded by the Deutsche Forschungsgemeinschaft (DFG, German Research Foundation) under Germany's Excellence Strategy - EXC-2094 - 390783311.

\section*{Data Availability}


The data underlying this article will be shared on reasonable requests to the corresponding author.



\bibliographystyle{mnras}
\bibliography{references} 




\appendix

\section{Embryo growth in the fixed-luminosity, $\beta{=}4$} 
\label{sec:appendix} 

As in Section \ref{sec:gromap}, we assume all pebbles to have a constant size of 1 mm (supported by mm-wavelength observations of disks \citep{Draine2006ONDISKS, Perez2015GRAIN25}).  We focus on this pebble size as it is the most efficient for pebble accretion \citep{Johnston2024FormationStars}.  In this regime, pebbles are all 1 mm but have a varying Stokes number depending on their location in the disk and the stage of disk evolution (Equations \ref{eq:peb_acc}- \ref{eq:pebh}).  At later stages and greater disk radii, the pebbles have larger Stokes number which results in them essentially becoming de-coupled from the movement of the gas and behaving like planetesimals.  Consequently, less efficient pebble accretion occurs \citep{Ormel2010TheDisks, Ormel2018CatchingPebbles}.  The initial mass of the embryo is calculated using Equation \ref{eq:embryo_mass}, via streaming instability.  The rest of the setup follows Table \ref{tab:pmig}. 

\begin{figure*}
    \includegraphics[width=\textwidth]{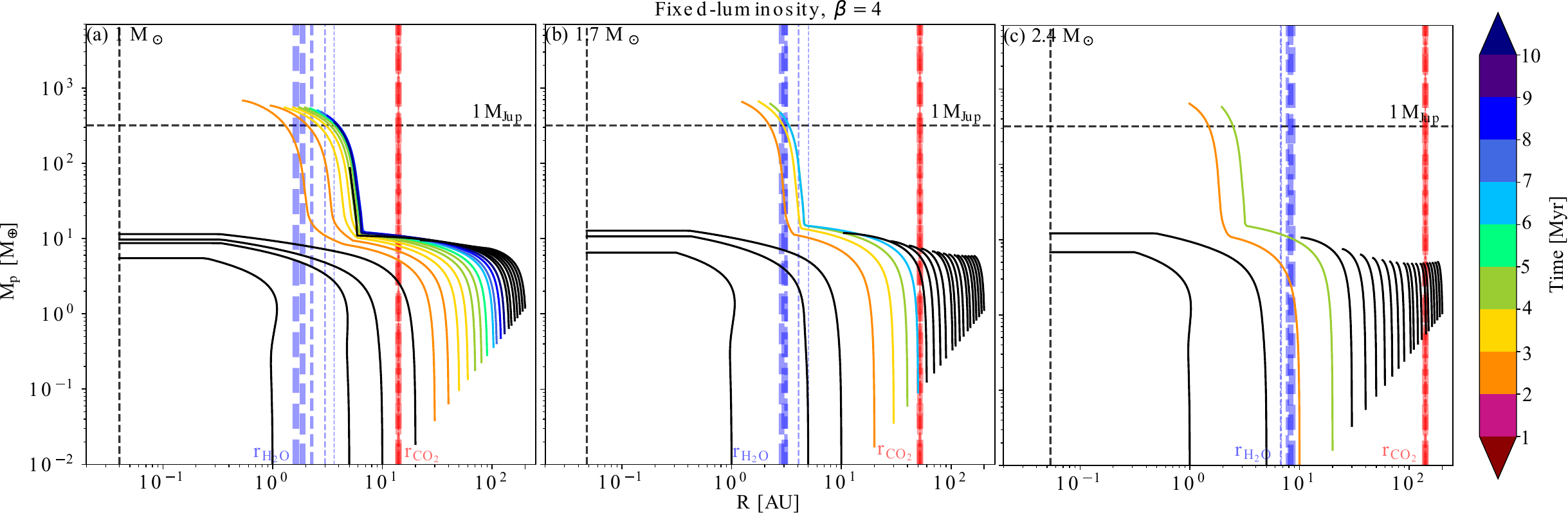}
    \caption{Migration and growth tracks of planets as functions of planet location $r_{0}$ and planet mass $M_{\rm p}$ around stars of (a) $1\ \rm M_{\odot}$, (b) $1.7\ \rm M_{\odot}$, and (c) $2.4\ \rm M_{\odot}$.  This is the fixed-luminosity case for $L_{\star} \propto M_{\star}^{\beta}$ when $\beta{=}4$.    The initial accretion rate is $\dot{M}_{\rm g0}{=}10^{-7} \Ms \rm yr^{-1}$.  The initial characteristic disk sizes are all $R_{\rm d0}{=}160$ AU.  The horizontal dashed line is located at $318 \Me$ or $1 \rm M_{\rm Jup}$ to denote our threshold minimum mass for a giant planet. The black vertical dashed line is the inner edge of our disk, the location where inward migration and planetary mass growth ceases (Equation \ref{eq:r_in}).  The blue and red dashed lines are the locations of the $\rm H_{2}O$ and $\rm CO_{2}$ snowlines for times 0.5, 1, 2, 3 Myr, the thickness of the line increasing with time accordingly.  The black tracks mean the planet does not surpass $1 \rm M_{\rm Jup}$.  The successful giant planets are coloured with respective to the time when they surpass $300 \Me$, and can be found in the colour bar on the right hand side.}
    \label{fig:pmig_beta4}
\end{figure*}

It is clear from the fixed luminosity $\beta{=}4$ approximation (Figure \ref{fig:pmig_beta4}) that a very high level of stellar luminosity is detrimental to planet formation processes overall (\textit{e.g.} large scale height (Equation \ref{eq:aspect}); inefficient core assembly (Equation \ref{eq:peb_acc_eff}); large isolation mass (Equation \ref{eq:m_iso}).  While it is unlikely that a PMS star would exhibit this amount of brightness for the entirety of its PMS phase (see Figure \ref{fig:lum}), it may be a useful approximation when considering planet formation around less massive stars prone to flares or outbursts (like T Tauri stars).  

When $\beta{=}4$, the disk is $\sim 10$ K warmer at all distances from the host star than the $\beta{=}2$ stellar luminosity approximation.  This is reflected in the $r_{\rm CO_{2}}$ snowline being located at increased radial distances up to $\sim 150$ AU.  The water snowline is also displaced to a greater radial extent, most predominately around the $2.4 \Ms$ star.  The composition of giant planets in this regime would be very water-poor, as all complex chemistry would be relegated to the gas phase other than silicates.  Such a high stellar luminosity also has a compounding impact on the scale height of both the pebbles and the disk itself.  This causes inefficient pebble growth and is particularly compounded at large orbital radii.  Embryos born at these distances also suffer from a slower rate of migration - hence, a long migration timescale - that would inhibit growth and migration.  Thus, embryos born at moderate distances from the host star $(r_{0} \sim 15{-}30$ AU) are able to balance the slower migration rate for enhanced pebble accretion and become massive in $t \sim 2{-}6$ Myr.


\bsp	
\label{lastpage}
\end{document}